\documentclass[twocolumn,aps,pra,floatfix,longbibliography,superscriptaddress]{revtex4-1}

\usepackage{hyperref}
\usepackage{amsmath}
\usepackage{amsfonts}
\usepackage{amsbsy}
\usepackage{xcolor}
\usepackage{soul}
\usepackage{graphicx}
\newcommand{\spvec}[1]{\ensuremath{\mathbf{#1}}}

\newcommand{\unitvec}[1]{\hat{\mathbf{{#1}}}}

\newcommand{\EQREF}[1]{Eq.~(\ref{#1})}
\newcommand{\bea}{\begin{eqnarray}}
\newcommand{\eea}{\end{eqnarray}}
\newcommand{\beq}{\begin{equation}}
\newcommand{\eeq}{\end{equation}}

\newcommand{\bE}{{\bf E}}

\newcommand{\pola}{\hat{\mathbf{e}}}
\newcommand{\rv}{{\bf r}}
\newcommand{\br}{{\bf r}}

\newcommand{\Dc}{{\cal D}}

\newcommand{\eo}{\epsilon_0}
\newcommand{\<}{\langle}
\renewcommand{\>}{\rangle}
\renewcommand{\(}{\left(}
\renewcommand{\)}{\right)}
\renewcommand{\[}{\left[}
\renewcommand{\]}{\right]}

\newcommand{\Pc}{\mathcal{P}}
\newcommand{\Pcv}{\boldsymbol{{\cal P}}}

\newcommand{\commentout}[1]{{}}

\newcommand{\cbE}{\boldsymbol{\mathbf{\cal E}}}

\usepackage{float}

\begin{document}
\title{Interaction of light with planar lattices of atoms: Reflection, transmission and cooperative magnetometry}
\date{\today}
\author{G. Facchinetti}
\affiliation{\'Ecole Normale Sup\'erieure de Cachan, 61 avenue du Pr\'esident Wilson, 94235 Cachan, France}
\author{J. Ruostekoski}
\affiliation{Department of Physics, Lancaster University, Lancaster, LA1 4YB, United Kingdom}

\begin{abstract}
We study strong light-mediated resonant dipole-dipole interactions in two-dimensional planar lattices of cold atoms. We provide a detailed analysis for the description
of the dipolar point emitter lattice plane as a ``super-atom'', whose response is similar to electromagnetically-induced transparency, but which exhibits an ultra-narrow collective
size-dependent subradiant resonance linewidth. The super-atom model provides intuitively simple descriptions for the spectral response of the array, including the
complete reflection, full transmission, narrow Fano resonances, and asymptotic expressions for the resonance linewidths of the collective eigenmodes. We propose a protocol
to transfer almost the entire radiative excitation to a single correlated subradiant eigenmode in a lattice and show that the medium obtained by stacked lattice arrays can form
a cooperative magnetometer. Such a magnetometer utilizes similar principles as magnetometers based on the electromagnetically-induced transparency. The accuracy of the cooperative magnetometer, however, is not limited by the single-atom resonance linewidth, but the much narrower collective linewidth that results from the strong dipole-dipole interactions. 

\end{abstract}

\maketitle

\section{Introduction}

Thin planar arrays of resonant emitters can couple strongly to incident electromagnetic fields, effectively making the transmission process cooperative; see, e.g., Refs.~\cite{FedotovEtAlPRL2010,LemoultPRL10,PapasimakisEtAlAPL2009,Jenkins17}. The phenomenon is generic and can appear for different types of scatterers with a considerable variation of sizes and frequencies of radiation, representing strong interactions that the electromagnetic field mediates between the scatterers. The resulting collective response can qualitatively also be described by arrays of dipolar point emitters ~\cite{JenkinsLineWidthNJP,CAIT,Jenkins17}, many cases even when the size of the scatterer is not negligible compared with the resonant wavelength. 

Atoms are ideal  dipolar point emitters and can be trapped in regular arrays using, e.g., optical lattice potentials~\cite{singlespin} or optical dipole traps~\cite{Browaeys_trap}. The idea that a two-dimensional (2D) planar array of atoms could be utilized for manipulation of light and for engineering a strong cooperative optical response with the associated super- and subradiant~\cite{Dicke54} excitation eigenmodes  was recently put forward~\cite{Jenkins2012a}.  Among the most dramatic properties of the planar arrays of dipolar point emitters with subwavelength spacing are ultra-sharp Fano resonances of transmission, with full transmission of light at the resonance and complete reflection at a range of other frequencies~\cite{CAIT}. Such systems formed by cold atoms could also exhibit quantum entanglement~\cite{Ritsch_subr}, be prepared to a giant subradiant excitation~\cite{Facchinetti} and states with a nontrivial topology~\cite{Perczel,Bettles_17topo}.
The research in cooperative light coupling with regular planar arrays has been expanding rapidly~\cite{Bettles_lattice,Yoo16,Bettles_prl16,Wang17,Genes17,Shahmoon,Asenjo_prx},
and with the models similar to those in Ref.~\cite{Olmos13} could be utilized to engineer also various interacting many-atom Hamiltonians.

Here we provide an expanded followup study of the recent work on dipolar point emitter arrays~\cite{CAIT,Facchinetti} where the cooperative response of a large array is modelled by introducing an effective \emph{super-atom} model, with only two coupled modes of radiative excitations. 
The super-atom model is reminiscent of the electromagnetically-induced transparency (EIT)~\cite{FleischhauerEtAlRMP2005} of independently scattering atoms.
The model predicts the response in the limit of an infinite array, and was shown in Refs.~\cite{CAIT,Facchinetti} to provide semi-analytic expressions for the conditions of complete reflection, full transmission, Fano resonances, and the resonance linewidths of the relevant  collective radiative excitation eigenmodes. For finite arrays we find that it provides a good qualitative agreement with the exact numerical results of the spectral profiles. Although, e.g., the total reflection can be demonstrated numerically (for recent studies, see also Refs.~\cite{Bettles_prl16,Shahmoon}), the super-atom model provides a simple intuitive description of the physical process, together with a model for the entire scattering spectrum.

The possibility to manipulate optical excitations was demonstrated in Ref.~\cite{Facchinetti} with a protocol to transfer excitations to a radiatively isolated collective state to form a giant subradiant excitation. The super-atom model is sufficient to describe the dynamics of the state transfer that is achieved by rotating the collective atomic polarization by an effective magnetic field. Here the two states of the super-atom represent the two orthogonal orientations of the dipoles in phase-coherent collective excitations. Depending on the size of the lattice and the confinement of the atoms, up to 98-99\% of the total excitation is found in a \emph{single} many-atom subradiant eigenmode. The prepared state is fundamentally different from the experimentally realized two-particle subradiant states of ions~\cite{DeVoe} or molecules~\cite{Hettich,McGuyer,Takasu}, or from collective states in atom clouds where a small fraction of atoms exhibit suppressed emission~\cite{Guerin_subr16}. Massive, spatially extended subradiance has only recently been experimentally observed in strongly coupled planar scatterer arrays~\cite{Jenkins17}.

Since resonant emitters play a key role in optical devices for classical and quantum technologies~\cite{HAM10}, and lattices of atoms are particularly relevant for precision measurements, e.g., in metrology~\cite{Nicholson_clock,Ye2016}, we apply the cooperatively coupled atomic arrays to precision measurements of weak magnetic fields. 
Unlike the conventional ideas of sensing~\cite{BudkerRamalisNatPhys2007} that are based on the independent scattering models, here the essential ingredient of the technique is the strong cooperative dipole-dipole interactions between the atoms, resulting in collective resonance linewidths and line shifts.
We consider strongly coupled arrays of atoms stacked together to form a 3D medium where the spatial separation between each planar lattice is sufficiently large (i.e., more than the optical wavelength), such that the interaction between the different lattices is weak. The basic idea is closely related to the EIT magnetometry, based on the sharp dispersion in a transparent media~\cite{Fleisch_magneto}.
However, as the EIT magnetometry is based on the independent atom scattering, the frequency scales for the suppression of absorption and the sharp dispersion are determined by the \emph{single-atom} resonance linewidth. In the proposed cooperative magnetometry the resonance linewidth of the relevant subradiant collective eigenmode can be dramatically narrower, leading to improved accuracy.

The paper is organized as follows. The basic model of the atom-light coupling is introduced in Sec.~\ref{sec:atomlight}. We begin by a single-atom description of the $J=0\rightarrow J'=1$ transition in the absence and presence of the Zeeman energy level shifts in Sec.~\ref{sec:singleatom}. This illustrates the basic principle of the rotation of the atomic polarization between two orthogonal directions that is then translated to the may-atom picture in Sec.~\ref{sec:manyatom}, where we introduced both the microscopic model of atom-light interactions and the effective super-atom model. The super-atom model is explored in detail in Sec.~\ref{sec:superatom}. We compare the effective model with the exact numerical simulations in Sec.~\ref{sec:comparison} and highlight the similarities of the model to the EIT of independent atoms in Sec.~\ref{sec:eit}. Complete reflection and full transmission, semi-analytic models for the collective resonance linewidths, pulse delays, and spectrum of the scattered light are analyzed in Sec.~\ref{sec:spectrum}. The cooperative magnetometry is described in Sec.~\ref{sec:magneto} and finally some concluding remarks are made in Sec.~\ref{sec:conc}.

\section{Atom-light coupling}
\label{sec:atomlight}

We consider a 2D square planar lattice of cold atoms in the $yz$ plane where the resonant dipole-dipole interactions are mediated between the atoms by the scattered light.
We assume that the atoms are illuminated by an incident weak-intensity laser propagating in the $x$ direction with the amplitude
\beq
\cbE(\rv)= {\cal E}_0(y,z) \pola_y \exp(ikx)\,,
\eeq
 with the linear polarization $\pola_y$ and the amplitude ${\cal E}_0(y,z)$ representing either constant or
a Gaussian profile on the $yz$ plane, perpendicular to the propagation direction.
Here, and in the rest of the paper, all the field amplitudes and the atomic polarization correspond to the slowly varying positive
frequency components with oscillations at the laser frequency $\omega$.
We consider a near-resonance $J=0\rightarrow J'=1$ atomic transition (Fig.~\ref{figure_level}) and assume a controllable Zeeman level
splitting of the $J'=1$ manifold. The Zeeman shifts could be induced
by magnetic fields or, e.g., by AC Stark shifts~\cite{gerbier_pra_2006}.

\subsection{Single-atom description}
\label{sec:singleatom}

\begin{figure}
	\centering
		\includegraphics[width=0.4\columnwidth]{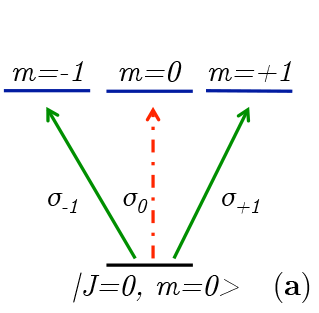}
	        \includegraphics[width=0.49\columnwidth]{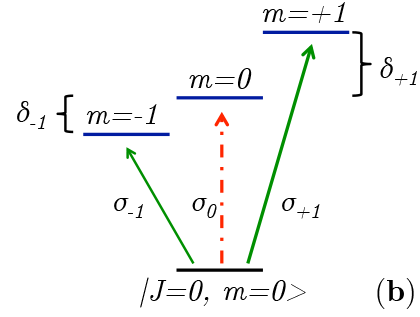}
    \vspace*{-11pt}
		\caption{The atomic level structure. The linearly polarized
		(along $y$) incident light propagates along the positive $x$ direction driving the $|J=0,m_J=0\>
		\rightarrow |J'=1,m_J=\pm1\>$ transitions. The light does not directly couple to the $|J=0,m_J=0\>
		\rightarrow |J'=1,m_J=0\>$ transition. The left panel shows the degenerate level structure and the right panel the
		level structure with the Zeeman splitting.
		 }
    \vspace*{-5pt}
		\label{figure_level}
\end{figure}

We first consider a single, isolated atom $j$.
The dipole moment of the atom reads
\beq
 {\bf d}_j= \Dc \sum_{\sigma} \pola_{\sigma} \Pc_{\sigma}^{(j)}\,,
 \eeq
where $\Dc$ denotes the reduced dipole matrix element.
The atom has three polarization amplitude components $\Pc_{\sigma}^{(j)}$ associated with the unit circular polarization vectors
\beq
\pola_{\pm1}=\mp {1\over \sqrt{2}} (\pola_x\pm i \pola_y), \quad \pola_0=\pola_z\,,
\eeq
that are coupled with the transitions
$|J=0,m=0\>\rightarrow |J'=1,m=\sigma\>$.
In the limit of low light intensity the excitation amplitudes satisfy
\beq
  \label{eq:dynamics_pol_single}
 \frac{d}{dt} \Pc^{(j)}_\sigma  = \left( i \Delta_\sigma - \gamma \right)
  \Pc^{(j)}_\sigma + i\frac{\xi}{{\cal D}} \unitvec{e}_\sigma^{\ast} \cdot
  \eo \cbE (\br_j) \textrm{,}
\eeq
where  $\xi=6\pi\gamma/k^3$ and the single-atom
Wigner-Weisskopf linewidth
\beq
\gamma={\Dc^2 k^3\over 6\pi\hbar\epsilon_0}\,.
\eeq
The detuning from the atomic resonance
$\Delta_{\sigma} = \omega-\omega_{\sigma}=\omega-(\omega_0+ \sigma\delta^{z}_\sigma)$ where $\omega_0$ is the resonance frequency of
the $|J=0\>\leftrightarrow |J'=1,m=0\>$ transition and $\pm\delta^{z}_{\pm}$ are the shifts of the $m=\pm1$ levels (Fig.~\ref{figure_level}).
The $y$-polarized light then drives the atomic
polarization components $ \Pc_{\pm 1}^{(j)}$ (Fig.~\ref{figure_level}). Here we instead write the equations of motion
in the Cartesian basis
\beq
{\bf d}_j = \Dc (\pola_{x} \Pc_{x}^{(j)} +\pola_{y} \Pc_{y}^{(j)} +\pola_{z} \Pc_{z}^{(j)}) \,,
\eeq
 such that
 \begin{align}
\Pc_{x}^{(j)} &={1\over \sqrt{2}} ( \Pc_{-1}^{(j)}- \Pc_{+1}^{(j)}),\\
\Pc_{y}^{(j)} &=-{i\over \sqrt{2}} ( \Pc_{-1}^{(j)}+ \Pc_{+1}^{(j)})\,.
\end{align}
We obtain
\begin{subequations}
\begin{align}
\dot\Pc^{(j)}_x & = (i \Delta_0-i\tilde\delta -\gamma) \Pc^{(j)}_x - \bar\delta  \Pc^{(j)}_y,\\
\dot \Pc^{(j)}_y & = (i \Delta_0-i\tilde\delta -\gamma) \Pc^{(j)}_y + \bar\delta  \Pc^{(j)}_x +i\xi\eo {\cal E}_0/\Dc\,,
\end{align}
\label{eq:singleatom}
\end{subequations}
where
\beq
\tilde\delta ={1\over 2} (\delta^{z}_{+}-\delta^{z}_{-}), \quad \bar\delta = {1\over 2} (\delta^{z}_{+}+\delta^{z}_{-})\,,
\label{eq:deltadef}
\eeq
 and $\Delta_0$ denotes the detuning
of the $m=0$ state. The incident light directly drives only
 $\Pc^{(j)}_y $, but the energy splitting of the levels $|m=\pm1\>$ introduces a coupling between  $\Pc^{(j)}_x $ and  $\Pc^{(j)}_y $. Although
the incident field is perpendicular to $\Pc^{(j)}_x $, the light can therefore still excite $\Pc^{(j)}_x $ by first driving $\Pc^{(j)}_y $. The $J=0\rightarrow J'=1$
 transition is isotropic when the excited-state energies are degenerate and any orientation of the orthogonal basis also forms an eigenbasis.
For $\bar\delta\neq0$,  $\Pc^{(j)}_{x/y}$ no longer are eigenstates. The dipoles are consequently turned toward the $x$ axis by the rotation around the effective magnetic field.

\subsection{Many-atom dynamics for a planar array}
\label{sec:manyatom}

\subsubsection{Microscopic model}

We next describe the response of the entire planar array where the light-mediated interactions between the atoms are taken into account. 
Analogous simulation methods can also be used in the studies of other resonant emitter systems, such as solid-state circuit resonators and plasmonics~\cite{JenkinsLongPRB,CAIT,Jenkins17}.
The single-atom picture
of the light excitation and the rotation of the atomic excitation can be translated to the many-atom language. The dynamics of the polarization  amplitudes in the many-atom picture
are obtained from the single-atom ones
\eqref{eq:dynamics_pol_single} by replacing the term representing the external field contribution~\cite{Ruostekoski1997a,Lee16}
\beq
  \label{eq:dynamics_pol}
 \frac{d}{dt} \Pc^{(j)}_\sigma  = \left( i \Delta_\sigma - \gamma \right)
  \Pc^{(j)}_\sigma + i\frac{\xi}{{\cal D}} \unitvec{e}_\sigma^{\ast} \cdot
  \eo \spvec{E}_{\rm ext}(\rv_j) \textrm{,}
\eeq
that could also be generalized to multi-level systems of alkali-metal atoms~\cite{Lee16,Jenkins_long16,Sutherland_satur}.
Here each amplitude in Eq.~\eqref{eq:dynamics_pol} in the atom $j$ is driven
by the sum of the incident field and the fields scattered from all the $N-1$ other atoms
\beq
  \label{eq:lightmedia}
\bE_{\rm ext}(\br_j) = \cbE (\br_j)+\sum_{l\neq j}
\bE^{(l)}_S(\br_j)\,.
\eeq
The scattered dipole radiation field from the atom
$l$ is
\beq
  \label{eq:scatteredlight}
\epsilon_0\bE^{(l)}_S(\br)={\sf G}({\bf r}-{\bf r}_l) \Dc \sum_{\sigma} \pola_{\sigma} \Pc_{\sigma}^{(l)}\,,
\eeq
 where
${\sf G}(\rv)$ is the dipole radiation kernel, such that $\bE^{(l)}_S(\br)$
represents the electric field at $\br$ from
a dipole $\Dc \sum_{\sigma} \pola_{\sigma} \Pc_{\sigma}^{(l)}$  residing at $\br_l$~\cite{Jackson}.

Owing to the resonant dipole-dipole interactions the atoms respond collectively to light, exhibiting collective
excitation eigenmodes with distinct collective radiative linewidths and line shifts.
The collective radiative excitation eigenmodes of the full system of $N$ atoms can be solved by representing the coupled system of atoms and light
as
\beq
\dot{{\bf b}} = i \mathcal{H}{\bf b} + {\bf F}\,,
\eeq
 where ${\bf b}$ is a vector made of the amplitudes $\Pc^{(j)}_\sigma$,
 \beq
{\bf b}_{3j-1+\sigma} = \Pc^{(j)}_\sigma\,,
 \eeq
with $\sigma=-1,0,1$ and $j=1,\ldots,N$. The driving of the dipoles by the incident light~\cite{JenkinsLongPRB} [due to the incident light field contribution from \EQREF{eq:lightmedia} in the last term in \EQREF{eq:dynamics_pol}] is given by
\beq
{\bf F}_{3j-1+\sigma}= i\frac{\xi}{{\cal D}} \unitvec{e}_\sigma^{\ast} \cdot  \eo   \cbE (\br_j)\,.
\eeq
The coupling matrix  $i\mathcal{H}$
provides the light-induced interactions between the atoms [due to the scattered light contribution from \EQREF{eq:lightmedia} in the last term in \EQREF{eq:dynamics_pol}] and the first terms on the right-hand-side of \EQREF{eq:dynamics_pol}, such that the diagonal terms read
\beq
\mathcal{H}_{3j-1+\sigma,3j-1+\sigma} =  \Delta_\sigma +i\gamma 
\eeq
and the off-diagonal terms (for $j\neq l$)
\beq
\mathcal{H}_{3j-1+\sigma,3l-1+\sigma'} =  i {\xi} \unitvec{e}_\sigma^{\ast} \cdot  {\sf G}({\bf r}_j-{\bf r}_l)   \pola_{\sigma'}
\eeq

The matrix $\mathcal{H}$ has $3N$ eigenmodes $\mathrm{v}_j$ with the eigenvalues $\delta_j + i \upsilon_j$
where $\delta_j=\omega_0-\omega_j$ is the shift of the collective mode resonance $\omega_j$
from the single atom resonance
and $\upsilon_j$ is the collective radiative linewidth.

\subsubsection{The effective super-atom model for the array}

We construct an effective super-atom model to describe the collective optical response of the planar array of atoms.
We will show that we can qualitatively understand the response by analyzing the behavior of the most dominant modes by formulating a super-atom model that only incorporates those modes. In this paper we consider incident field profiles that are smoothly varying in the perpendicular direction to the propagation direction of light.
The incident light is then phase-matched to a smoothly-varying, phase-coherent excitation of the atoms. Note that this situation is quite different from the planar array excitations considered in Ref.~\cite{Jenkins2012a} where the phase profile of the incident field was modulated. This modulation resulted in spatially-localized excitations.

\begin{figure}
	\centering
		\includegraphics[width=0.65\columnwidth]{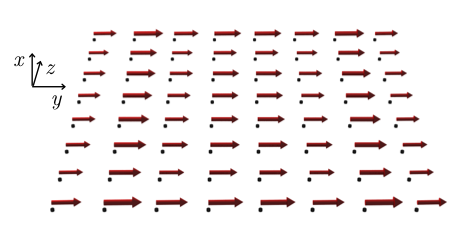}
	        \includegraphics[width=0.65\columnwidth]{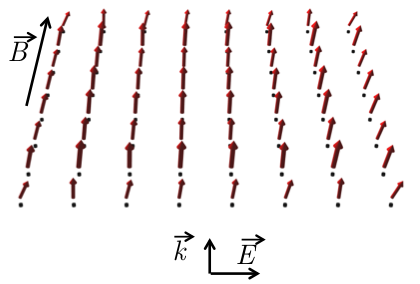}
	        \includegraphics[width=0.65\columnwidth]{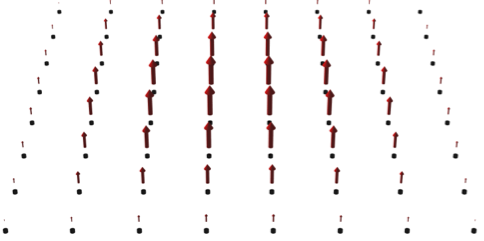}
    \vspace*{-5pt}
		\caption{
		Illustration of the phase-coherent response of the atoms, displaying the numerically calculated (based on the microscopic model) steady-state solutions to a small square 2D  $8\times 8$ array with one atom per site and the lattice spacing $a=0.55\lambda$.
		The linearly polarized (along $y$) incident plane-wave light field propagates along the positive $x$ direction.
		On the top panel the Zeeman shifts vanish and the atomic dipoles oscillate in the array plane. In the middle panel the effective Zeeman splitting (due to a real or synthetic magnetic field along the $z$ axis) drives the dipoles
		to be oriented normal to the plane with $(\delta^{z}_{+},\delta^{z}_{-}) = (0.45,1.75)\gamma$. The bottom panel shows the collective eigenmode of the coupled system of light and atoms in the absence of the Zeeman shifts that
		closely resembles the uniform excitation of the atomic dipoles normal to the array plane. The dipoles oscillate normal to the planar array and the excitation amplitude decays close to the edge of the lattice.
 }
    \vspace*{-11pt}
		\label{figure_lattice}
\end{figure}

It turns out that in the common situations that we are interested here we only need to consider two collective excitation eigenmodes (the eigenmodes of the collective system of atoms and light that are evaluated in the absence of the Zeeman shifts).
The linear polarization couples to a collective (``coherent in-plane'') mode in which all the dipoles are coherently oscillating along the $y$ direction with the excitation $ \Pc_I$; see Fig.~\ref{figure_lattice}. When we introduce nonvanishing Zeeman shifts, this mode is no longer an eigenmode. The situation is now similar to the single-atom case where the atomic dipole is turned toward the $x$ axis due to the Zeeman splitting. In the collective lattice system that is driven by the laser with a uniform phase profile along the lattice, the atomic dipoles keep oscillating in phase even when the excited state degeneracy is broken. The dynamics in the presence of the Zeeman shifts can therefore be qualitatively analyzed by a simple two-mode model when we assume that
$ \Pc_I$ is predominantly coupled with a phase-coherent collective (``coherent perpendicular'') excitation $ \Pc_P$ where all
the atomic dipoles are oscillating in phase, normal to the plane; see Fig.~\ref{figure_lattice}.  Also this mode is a collective eigenmode in the absence of the Zeeman energy splitting of the electronic excited state.
We can now establish an effective two-mode dynamics
\begin{subequations}
\begin{align}
\dot\Pc_P & = (i \Delta_P-i\tilde\delta -\upsilon_P) \Pc_P - \bar\delta  \Pc_I, \label{coll2modex}\\
\dot \Pc_I & = (i \Delta_I-i\tilde\delta -\upsilon_I) \Pc_I +  \bar\delta  \Pc_P +i\xi\eo {\cal E}_0/\Dc\,,
\label{coll2modey}
\end{align} \label{bothtwo}
\end{subequations}
where $\tilde\delta$ and $\bar\delta$ are given by \EQREF{eq:deltadef}, $\upsilon_{P}$ and $\upsilon_{I}$ are the collective linewidths of the corresponding eigenmodes of the many-atom system (for $\delta^{z}_{\pm}=0$) and
\beq
\Delta_{P/I}=\omega-\omega_{P/I}=\Delta_0+\delta_{P/I}
\eeq
are the detunings of the incident light from the resonances of these modes (that are shifted by the collective line shifts $\delta_{P}$ and $\delta_{I}$).

The coupled equations for the two phase-coherent collective modes in \EQREF{bothtwo} are almost identical to the equations describing the single-atom dynamics in the Cartesian coordinate system in \EQREF{eq:singleatom}. The only difference is
that the single-atom resonance linewidth $\gamma$ is replaced by the collective eigenmode resonance linewidths $\upsilon_{P}$ and $\upsilon_{I}$, and that the resonance of the two modes also exhibit the additional collective shifts $\delta_{P}$ and $\delta_{I}$.
The effective contribution of the collective light-mediated resonant dipole-dipole interactions is encapsulated in these collective linewidths and line shifts. The $3N$ equations for the $N$ atoms with three polarization components for the excitations are now replaced by two equations for the collective eigenmodes: one for the dipoles oscillating in the lattice plane and another one for the dipoles oscillating parallel to the propagation direction of the light and normal to the lattice plane.
The simplicity of the effective super-atom approach becomes obvious when we compare it with the full microscopic model in \EQREF{eq:dynamics_pol}: In the microscopic model for each atom we need to calculate the scattered fields from all the other atoms in $\spvec{E}_{\rm ext}(\rv_j) $.

The physical characteristics of the two eigenmodes are entirely different~\cite{Facchinetti}. Since all the dipoles in the collective mode $ \Pc_I$ are in the lattice plane,
$ \Pc_I$ is responsible for strong reflection and transmission of light that follows from the dipole radiation pattern directed out of the plane. However, the collective mode excitation $ \Pc_P$ dominantly radiates within the plane,
where the emission of light is directed toward the other atoms, enhancing light-mediated interactions between the atoms: For light to escape, it generally undergoes many scattering events. This is the physical explanation of the giant
subradiance demonstrated for the mode $ \Pc_P$ in Ref.~\cite{Facchinetti}. With the targeted excitation that is resonant with $ \Pc_P$ almost all the excitation can be driven to the mode $ \Pc_P$ by utilizing the Zeeman shifts in the rotation
of the atomic polarization normal to the plane~\cite{Facchinetti}.

\section{Simple description of the lattice by super-atom}
\label{sec:superatom}

\subsection{Comparisons with the exact calculations}
\label{sec:comparison}

\begin{figure}
	\centering
		\includegraphics[width=0.97\columnwidth]{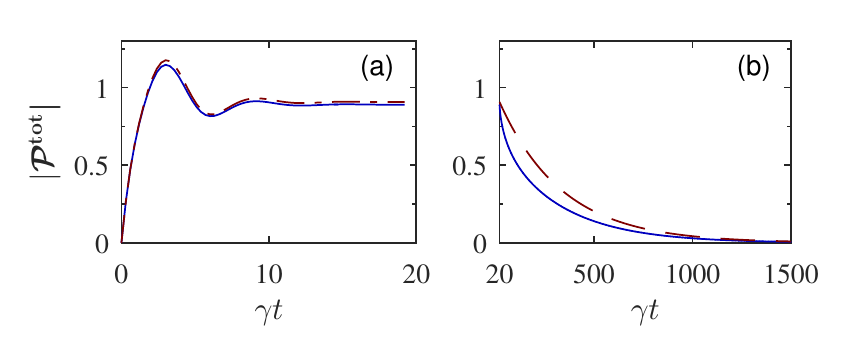}
		\includegraphics[width=0.97\columnwidth]{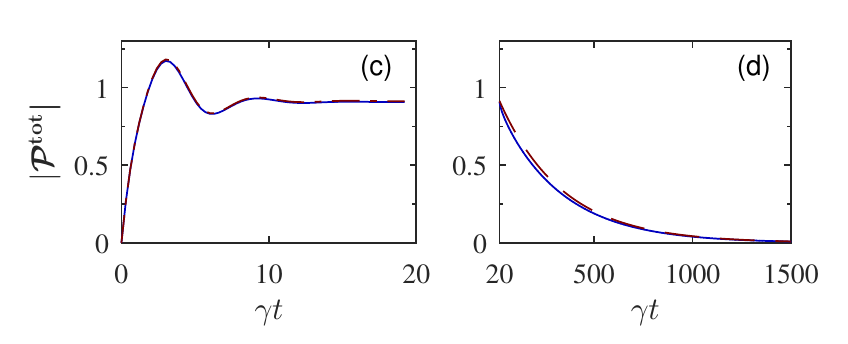}
    \vspace*{-12pt}
		\caption{Comparisons between the phenomenological super-atom model (dashed, red lines) and the exact numerical simulation (solid, blue lines). The Zeeman shifts and the detuning $(\delta_{+}^{z}, \delta_{-}^{z},\Delta_0)=(1.1,1.1,0.65)\gamma$ have been chosen such that the collective polarization in the steady-state response is pointing normal to the lattice plane. In the two-mode model we have used the numerical values of the full eigenvalue calculation  that are $\delta_P = -0.65\gamma$, $\delta_I = -0.68\gamma$, $\upsilon_P = 0.0031\gamma$, $\upsilon_I=0.79\gamma$. The initial evolution of the laser-driven lattice before reaching the steady-state distribution for the case of (a) plane-wave excitation, (c) Gaussian beam excitation. The evolution after the incident light and the Zeeman shifts are turned off for the case of (b) plane-wave excitation, (d) Gaussian beam excitation. In the case of a plane-wave excitation the super-atom model
		differs from the full numerical solution at early times due to the contribution of additional collective modes in the dynamics (see Fig.~\ref{fig:popu}). The additional modes decay faster and the dynamics of the calculations is more similar at later times when  the slowly decaying subradiant coherent perpendicular mode dominates.
 }
		\label{fig:2modevsexact}
\end{figure}

The two-mode model qualitatively captures many of the essential features of the full many-body dynamics, when we substitute the numerically calculated values of $\upsilon_{P/I}$ and $\delta_{P/I}$ of the colective eigenmodes. In Fig.~\ref{fig:2modevsexact} we show the comparison between the dynamics given by the two-mode model and the full numerics of all the 1200 collective excitation eigenmodes of the 20$\times$20 array for the lattice spacing $a=0.55\lambda$. In the numerical simulations we calculate the optical response by following the lattice simulation procedure introduced in Ref.~\cite{Jenkins2012a} by solving the microscopic model described by Eqs.~\eqref{eq:dynamics_pol},  \eqref{eq:lightmedia}, and~\eqref{eq:scatteredlight}.  In the limit of low light intensity, for stationary atoms with the $J=0\rightarrow J'=1$ transition the results are exact~\cite{Javanainen1999a,Lee16},
incorporating recurrent scattering between the atoms and light-induced correlations. In a random medium such position-dependent correlations can lead to a violation of standard continuous medium electrodynamics~\cite{Javanainen2014a,JavanainenMFT}.

We show in Fig.~\ref{fig:2modevsexact} the dynamics of the total polarization of the system $ |\Pcv^{\rm tot}| = | \sum_{j,k} \Pc^{(j)}_k \pola_k|/N$ [in all the numerical results, the polarization amplitudes are expressed
in the dimensionless form $\Pc\rightarrow \Dc \Pc k^3/(6\pi \eo {\cal E}_0)$].
The incident light [in Fig.~\ref{fig:2modevsexact}(a,b) a plane wave and in (c,d) a Gaussian beam] excites the $y$ components of the atomic dipoles.
The Zeeman shifts turn the polarization density toward the $x$ direction. At the resonance $(\delta^{z}_{+},\delta^{z}_{-},\Delta_0) = (1.1,1.1,0.65)\gamma$ we find the dipoles almost entirely
along the $x$ direction

The first part of the dynamics  [in Fig.~\ref{fig:2modevsexact}(a,c)] describes the excitation of the atomic dipoles before the system reaches the steady state configuration. Once in the steady state, the Zeeman shifts and the incident laser are
turned off, resulting in a decay of the excitations  [in Fig.~\ref{fig:2modevsexact}(b,d)]. Since the atomic dipoles here are approximately perpendicular to the plane, the excitation represents giant subradiance~\cite{Facchinetti} corresponding to a slow decay.
For the plane-wave excitation the decay rates of the two cases differ at early times, while in the case of a Gaussian incident field excitation the agreement is better.

\begin{figure*}
	\centering
		\includegraphics[width=1.5\columnwidth]{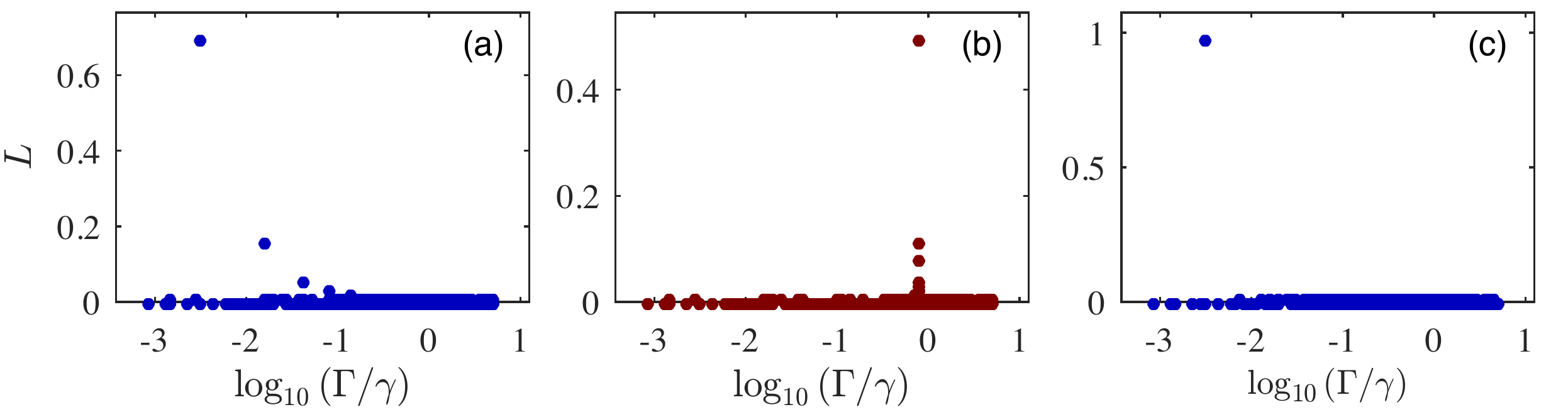}
    \vspace*{-8pt}
		\caption{The measure of the mode population $L$ of the different eigenmodes in the steady-state response for fixed atomic positions,
		(a) incident plane wave excitation, $(\delta^z_{+}, \delta^z_{-},\Delta_0) = (1.1,1.1,0.65)\gamma$; (b) incident plane wave excitation, $(\delta^z_{+}, \delta^z_{-},\Delta_0) = (0,0,0)\gamma$; (c) Gaussian beam excitation with the standard deviation $6a$, $(\delta^z_{+}, \delta^z_{-},\Delta_0) = (1.1,1.1,0.65)\gamma$. The eigenmodes are ordered by their collective radiative resonance linewidths on a logarithmic scale.
		In (b) the atomic polarization density is in the lattice plane and the excitation is dominated
		(over 50\% of the total excitation) by the collective mode where the dipoles are coherently oscillating in the $y$ direction. In (a) and (c) the atomic polarization density is pointing normal to the lattice plane,
		and the excitation is dominated by about (a) 70\%, (c) 98\% by the collective subradiant mode where the atomic dipoles are coherently oscillating in the $x$ direction. In (a) also some modes with notably
		broader resonances are occupied that manifests themselves in a faster initial decay of the radiative excitation; see Fig.~\ref{fig:2modevsexact}(b).
    }
    \vspace*{-11pt}
		\label{fig:popu}
\end{figure*}
In order to explain the differences between the two excitation methods and the accuracy of the super-atom approach, we show in Fig.~\ref{fig:popu} the populations of the different eigenmodes of $\mathcal{H}$ in the steady-state response.
Since $\mathcal{H}$  is non-Hermitian, we define an overlap measure \footnote{For the isotropic $J=0\rightarrow J'=1$ system $\mathcal{H}$ is
symmetric, and we can determine the biorthogonality condition $\mathrm{v}_j^T \mathrm{v}_i=\delta_{ji}$, except for possible zero-binorm states for
which $\mathrm{v}_j^T \mathrm{v}_j=0$ (that we have not encountered in our system).}
\beq
L_j= {| \mathrm{v}_j^T \mathrm{b}|^2\over \sum_i | \mathrm{v}_i^T  \mathrm{b} |^2}\,,
\label{eq:measure}
\eeq
for the eigenvector $\mathrm{v}_j$ in the state $ \mathrm{b}$. We also use this measure to determine the eigenmodes that are the closest to the ideal phase-coherent
polarization excitations $ \Pc_I$ and $ \Pc_P$.

In Fig.~\ref{fig:popu} we show the eigenmode populations for the cases where the subradiant mode with the polarization
vectors normal to the lattice array are excited using a plane wave and a Gaussian beam. For comparison, we also display the collective eigenmode populations when the steady-state dipoles are not rotated by the Zeeman shifts and they are in the array plane. In all the cases only a small number of modes are significantly excited. In the case of  driving of the dipole excitation normal to the lattice using a plane wave the excited eigenmodes have different linewidths. For the Gaussian beam driving, only one eigenmode is notably excited. This is because the Gaussian beam intensity is better matched with the density distribution of the eigenmode in Fig.~\ref{figure_lattice}.

The occupation of more than a single eigenmode with different decay rates in Fig.~\ref{fig:popu}(a) illustrates why the super-atom model for the plane wave excitation in Fig.~\ref{fig:2modevsexact}(b) differs at early times. In fact, one can show~\cite{Facchinetti}
that the fitting of the radiative decay to a double-exponential provides  a much better result than the fitting to a single exponential.
Although the slowly-decaying case is dominated by the subradiant $ \Pc_P$ excitation eigenmode  with about 70\% of the total excitation (and with the linewidth of $\upsilon_P \simeq 3.1\times 10^{-3}\gamma$),  the reason for the double-exponential decay is a prominent
excitation $\sim15\%$ of an additional eigenmode whose linewidth $\simeq0.015\gamma$ notably differs from that of $ \Pc_P$.

In the case of a Gaussian incident field excitation in Fig.~\ref{fig:popu}(c), the agreement between the exact solution and the super-atom model is better, since in that case the entire excitation is dominated by a single collective eigenmode in Fig.~\ref{fig:popu}(c) that can reach about 98-99\% of the total excitation~\cite{Facchinetti}. The decay rate in that case can also be better described by a single exponential. In case that the atomic polarization density is in the lattice plane in Fig.~\ref{fig:popu}(b), the excitation is dominated by the collective mode where the dipoles are coherently oscillating in the $y$ direction, and the decay rates of all the occupied modes are comparable to the linewidth of a single isolated atom.

\subsection{Analogy to electromagnetically-induced transparency}
\label{sec:eit}

\begin{figure}
	\centering
		\includegraphics[width=0.45\columnwidth]{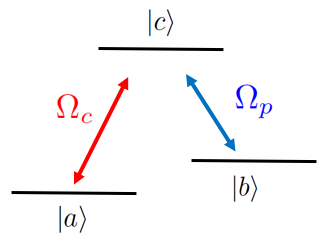}
		\includegraphics[width=0.45\columnwidth]{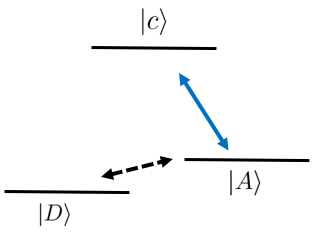}
    \vspace*{-11pt}
		\caption{Schematic illustration of the electromagnetically-induced transparency. The two electronic ground states  $|a\>$ and $|b\>$ are driven by the fields with the Rabi frequencies $\Omega_c$ (coupling)  and $\Omega_p$ (probe), respectively, to the same electronically excited state $|c\>$.
		The destructive interference between the transition amplitudes from the two electronic ground levels leads to
		the dark state $|D\>\propto \Omega_c |b\> - \Omega_p |a\>$ that is decoupled from the driving light. At the same time, the absorbing state $|A\> \propto  \Omega_p |b\> + \Omega_c |a\>$ is driven by the incident field. Small perturbations from the dark state result in coherent driving between the ground states.
		 }
    \vspace*{-5pt}
		\label{figure_eit}
\end{figure}

The effective super-atom model has an analogy with EIT~\cite{FleischhauerEtAlRMP2005}. In a typical EIT setup of a lambda-three-level atom, shown in Fig.~\ref{figure_eit}, the transition amplitudes from two electronic ground states $|a\>$ and $|b\>$ to the same electronic excited state $|c\>$ destructively interfere. The interference results in a spectral transparency window in an otherwise opaque medium where the absorption is suppressed. When the population of the excited state can be neglected, EIT has a classical mechanical analogy of coupled oscillating springs~\cite{eit_mech}. In the limit of low light intensity, the EIT equations for a single atom polarizations for the transitions from the `dark' $|D\>$ and `absorbing' $|A\>$ states, $|D\>\leftrightarrow |c\>$ and $|A\>\leftrightarrow |c\>$ , are reminiscent of those of the super-atom model for $\Pc_P$ and $\Pc_I$, respectively [Eq.~\eqref{bothtwo}]. Here the strongly subradiant collective eigenmode of the interacting \emph{many-atom} system where the dipoles are pointing normal to the plane plays the role of the \emph{single-particle} dark state superposition of the two ground states of the EIT system. The collective eigenmode where the dipoles are in the lattice plane represents the absorbing state superposition in EIT.

In EIT small perturbations from the dark state result in coherent driving between the ground states~\cite{DuttonHau,dutton_prl_2004}. This was utilized in the stopping and storage of light pulses by controlling the atom
population by means of turning on and off the coupling field~\cite{LiuEtAlNature2001}. In the subradiant state preparation~\cite{Facchinetti}  the idea of transferring almost the entire atom population into a single eigenmode is based on an analogous principle: the atoms are coherently driven to the subradiant mode with the Zeeman shifts that are subsequently turned off after the system has reached the steady state configuration.

In the following sections the analogy between the super-atom for the lattice and single-particle EIT is further illustrated by examples of reflection, transmission, and Fano resonances. We also propose a magnetometer based on cooperative light-induced interactions that has similar principles as the EIT-magnetometry.

\subsection{Full reflection and transmission for the array}
\label{sec:spectrum}

The planar arrays of resonant emitters have been demonstrated as powerful tools for manipulating electromagnetic radiation. The cooperative resonance phenomena of regular planar arrays consisting of dipolar point scatterers were analyzed
in detail in Ref.~\cite{CAIT}, where simple arguments for the complete reflection and transmission, and the emergence of Fano resonances were developed. The complete reflection for planar arrays of atomic dipoles for different geometries was also studied in Ref.~\cite{Bettles_prl16}, and the analysis of Ref.~\cite{CAIT} was further extended in Ref.~\cite{Facchinetti}.

Here we summarize the super-atom analysis of the transmission spectrum of Refs.~\cite{CAIT,Facchinetti} with the focus on the complete reflection and transmission limits. These results are then used in the magnetometry example based on the stacked arrays of atomic dipoles in Sec.~\ref{sec:magneto}.

We express the reflectance and the transmittance amplitudes in terms of the incident and the scattered field amplitudes $ \spvec{E}_I$ and $ \spvec{E}_S$, respectively
  \beq
  	  r = \frac{\unitvec{d}\cdot
      \spvec{E}_S(-\unitvec{e}_x)}{ \unitvec{d} \cdot \spvec{E}_I(\unitvec{e}_x)},\quad
          t = \frac{\unitvec{d}\cdot\left[\spvec{E}_I(\unitvec{e}_x) +
        \spvec{E}_S(\unitvec{e}_x)\right]}{ \unitvec{d} \cdot
      \spvec{E}_I(\unitvec{e}_x)} \textrm{,}
      \label{eq:reftra}
  \eeq
  where $\spvec{E}_S(\unitvec{n})$, with $\unitvec{n}=\pm \unitvec{e}_x$,  describes the scattered light to the forward/backward direction from the lattice in the $yz$ plane.
The reflection and transmission amplitudes can be numerically calculated by solving for the optical response of the entire array of atoms using the microscopic model described by Eqs.~\eqref{eq:dynamics_pol},  \eqref{eq:lightmedia}, and~\eqref{eq:scatteredlight}. The description,
however, becomes particularly simple with the super-atom model, even without the precise knowledge of the values of the collective eigenmode resonance linewidths and line shifts, $\upsilon_{P}$, $\upsilon_{I}$, $\delta_{P}$, and $\delta_{I}$.

The steady-state solution of the super-atom model of Eq.~\eqref{bothtwo} is easily obtained
\begin{subequations}
\begin{align}
  \label{eq:ssI}
 	\Pc_I &= -i\frac{Z_P(\Delta_0)}{\bar{\delta}^{2} - Z_P(\Delta_0)Z_I(\Delta_0)}f \\
	\label{eq:ssIb}
 	\Pc_P &= -i\frac{\bar{\delta}}{Z_P(\Delta_0)}\Pc_I \textrm{,}
\end{align}
\end{subequations}
where we have introduced the abbreviation $f=i\xi\eo {\cal E}_0/\Dc$, and defined
\begin{align}
  \label{eq:ZDef}
  Z_{P/I}(\Delta_0) & \equiv  \Delta_{P/I}  -\tilde{\delta} + i \upsilon_{P/I} ,\\
  \Delta_{P/I} & \equiv \Delta_0 + \delta_{P/I}\,.
\end{align}

\subsubsection{Complete reflection}

The atomic dipoles in the collective eigenmode excitation $ \Pc_P$ are oriented in the direction normal to the plane and cannot emit in the exact forward or back direction.
Therefore the scattering in the exact forward and back direction is solely produced by the coherent in-plane collective mode $ \Pc_I$.
Consequently, we obtain the steady-state reflection from the steady-state solution to $\Pc_I$ amplitudes, given by Eq.~\eqref{eq:ssI},
  \begin{equation}
  \label{reflect1}
    r = - r_0\, \frac{i\upsilon_I Z_P(\Delta_0)}{\bar{\delta}^{2} - Z_P(\Delta_0)Z_I(\Delta_0)} \,,
\end{equation}
where $r_0$ denotes the reflectance amplitude at the resonance of the coherent in-plane collective mode ($Z_I=i\upsilon_I$) when the Zeeman shifts vanish $\bar{\delta}=0$.
We also have
the symmetry $\spvec{E}_S(\unitvec{e}_x)=\spvec{E}_S(-\unitvec{e}_x)$ in Eq.~\eqref{eq:reftra}, resulting in
\beq
t=1+r\,.
\label{symmetrybf}
\eeq

The reflection and transmission properties of a planar lattice of atoms are described in the super-atom model by the simple relations Eqs.~\eqref{reflect1} and~\eqref{symmetrybf}.
The scattering problem can be further simplified when we consider an infinite lattice on the $yz$ plane. The 2D lattice behaves similarly to a 2D diffraction grating where each atom emits light. The lattice structure generates the diffraction pattern, while the atomic wave functions on individual sites correspond to the Debye-Waller factors and are responsible for overall envelope of the diffraction pattern~\cite{Ruostekoski09,Elliottt15}. For subwavelength lattice spacing ($a<\lambda$) only the zeroth order Bragg peak of the scattered light survives in the far field. This corresponds to the exact forward and back scattered light. Since we defined the reflection and transmission amplitudes according to the scattered light in \EQREF{eq:reftra}  in the exact forward and back directions, the energy conservation  then states that
\beq
|r|^2+|t|^2=1\,.
\eeq
Combining this with Eq.~\eqref{symmetrybf}  yields
\beq
-{\rm Re}(r)=|r|^2\,,
\label{constrainteq}
\eeq
whose solutions generally are complex, but
all the \emph{real} solutions are $r=-1$ and 0, corresponding to perfect reflection and perfect transmission, respectively. With the vanishing Zeeman shifts ($\bar\delta=0$)~\footnote{In this limit we only have an equation for $ \Pc_I$, since there is no coupling to $ \Pc_P$ in Eq.~\eqref{bothtwo}.}, Eq.~\eqref{reflect1}  reduces to
 \begin{equation}
    r = r_0\, \frac{i\upsilon_I }{ Z_I(\Delta_0)} \,.
\end{equation}
Even without any additional knowledge of the scattering properties of the system, we can take $r_0$ to be real by appropriately redefining the resonance frequencies.
(The scattering calculation yields a real-valued $r_0$ even without  a redefined resonance, as shown below.)
The only value corresponding to a nonvanishing reflection is then
\beq
r_0=-1,
\label{eq:total}
\eeq
indicating that an incident field at the resonance of the coherent in-plane collective mode experiences a total reflection when the Zeeman shifts are zero.
Note that the amplitudes of the scattered light are equal in the forward and back directions, but in the forward direction the scattered light is precisely cancelled by interference with the incident field.

The expression for $r_0$ may, in fact, be also derived independently by explicitly evaluating the scattered light. In Ref.~\cite{CAIT} the response was analyzed by calculating the transmission amplitude of \EQREF{eq:reftra} by deriving the ratio between the light amplitudes using the Fraunhofer diffraction theory from a rectangular aperture. Taking then the limit of an infinite lattice yielded the result \footnote{Owing to the different definition of the linewidth in Ref.~\cite{CAIT}, the atomic lattice result is obtained with the substitution $\Gamma_E\rightarrow \gamma/2$ in the expression of the meta-molecule linewidth.}
\beq
r_0= - {3\gamma\lambda^2\over 4\pi  \upsilon_I a^2}\,.
\eeq
Combining this with \EQREF{eq:total} yields an expression for the linewidth of the collective eigenmode $\upsilon_I$ where the dipoles are oscillating in the plane of the array~\cite{CAIT}
\beq
\lim_{N\rightarrow \infty} \upsilon_I = {3\lambda^2 \over 4\pi a^2} \gamma\,.
\label{eq:analyticline1}
\eeq
The result was shown to agree with the numerical simulations in the large lattice limit in Ref.~\cite{CAIT}.
The assumption in the derivation that the lattice spacing is less than the wavelength, $a<\lambda$, sets the minimum limit how subradiant the linewidth of this mode can be $\upsilon^{\rm min}_I /\gamma= {3 /( 4\pi)} $. Moreover, when $a< \sqrt{3/\pi}\lambda/2$, the mode becomes super-radiant in the infinite lattice limit.

Although similar results to those of Eqs.~\eqref{eq:total} and~\eqref{eq:analyticline1} were also obtained using different methods in Ref.~\cite{Shahmoon}, the original derivations for the complete reflection that we have described here were already presented in Refs.~\cite{CAIT,Facchinetti}.

We can also solve the full expression \eqref{reflect1} for the reflection amplitude with the constraint \eqref{constrainteq}. Since we have $r_0\neq0$ and $\upsilon_I\neq0$, these can be solved for arbitrary $\bar\delta$ only for $\upsilon_P=0$. We obtain another result for the collective eigenmode resonance linewidths,
\beq
\lim_{N\rightarrow \infty} \upsilon_P = 0\,.
\label{eq:analyticline2}
\eeq
In the limit of an infinite lattice, the subradiant mode where the dipoles oscillate perpendicular to the lattice becomes entirely decoupled from light with the zero linewidth. We have also numerically demonstrated it: For fixed atomic positions the mode becomes increasingly more subradiant in larger lattices with $\upsilon_P/\gamma \simeq N^{-0.91}$~\cite{Facchinetti}.

\subsubsection{Complete transmission}

In the previous section we analyzed the response of the system when driven at the resonance of the $\Pc_I$ mode. We will now simplify the expression \eqref{reflect1} for the reflection amplitude at the resonance of the $\Pc_P$ mode, i.e., when $\Delta_P = \tilde{\delta} $. We find
\begin{equation}
  \label{eq:R_delta_M}
  r(\Delta_P = \tilde{\delta}) \approx - \frac{\upsilon_I\upsilon_P}{\bar{\delta}^2 + \upsilon_I\upsilon_P}
\end{equation}
where we have assumed $|\delta_P - \delta_I |\ll \upsilon_I$
so that we can neglect any difference between the $\Pc_I$ and
$\Pc_P$ resonance frequencies. (We find that this holds approximately true for the lattice $a\simeq 0.55\lambda$.)
When the Zeeman shifts satisfy
\beq
\bar{\delta}^{2} \gg \upsilon_P\upsilon_I\,,
\label{eq:transmission}
\eeq
reflectance on
$\Pc_P$ resonance is therefore suppressed, and transmittance is enhanced.
Remarkably, in the limit of a large array the response can therefore vary between a complete reflection and full transmission.

\subsubsection{Spectrum of the scattered light}

In the previous sections we found that in the large lattice limit the optical response can vary between the complete reflection ($\bar\delta=0$) and full transmission ($ \Pc_P$ resonance). We will now analyze the entire spectrum of scattered light that can exhibit very narrow features, related to the Fano resonances.
The narrow linewidth $\upsilon_{P}$ of the subradiant eigenmode manifests itself  in these resonances.

\begin{figure*}
	\centering
    \vspace*{11pt}
		\includegraphics[width=1.7\columnwidth]{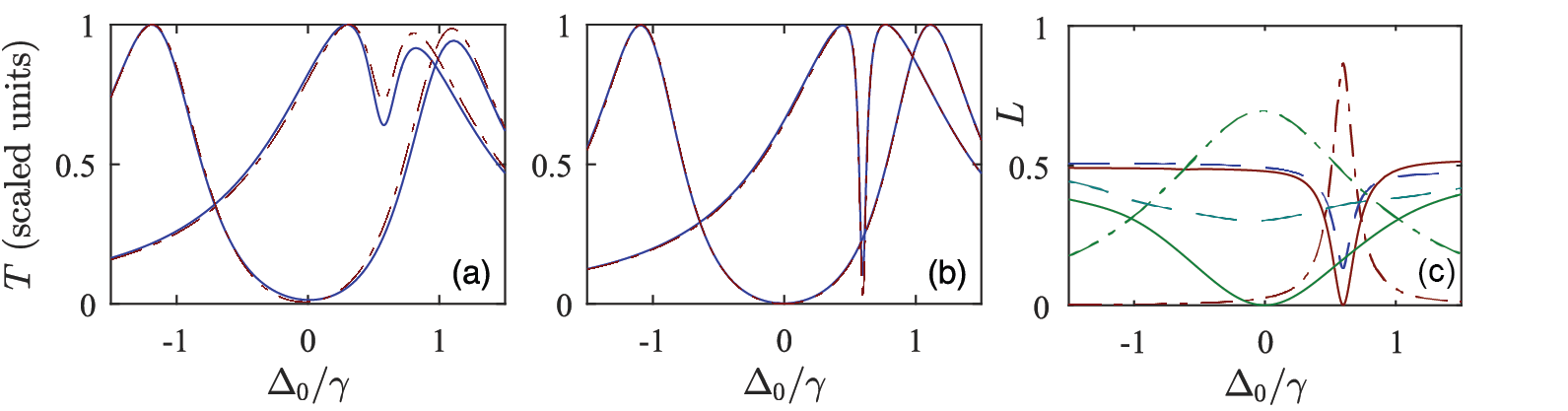}.
    \vspace*{-10pt}
		\caption{The spectrum of forward or back scattered light for the lattices of (a) 3$\times$3; (b) 20$\times$20 sites, and (c) the eigenmode populations $L$  corresponding to the spectral curves of (b).
		The curves are shown in the steady-state response  as a function of the frequency of the incident light, and the spectrum of the larger lattice in (b) displays a sharp Fano resonance.
		We show in (a,b) both the super-atom model [red, dashed curves; with numerically calculated eigenvalues for the two dominant eigenmodes $(\delta_P +i\upsilon_P)/\gamma\simeq
-0.65+0.0031i$ (20$\times$20), $-0.62+0.12i$ (3$\times$3),  and $(\delta_I+i\upsilon_I)/\gamma\simeq
-0.68+0.79i$ (20$\times$20), $-0.59+0.83i$ (3$\times$3)]  and the exact numerical results (blue, solid curves). The two cases in (a,b) represent the two different orientations
of the atomic dipoles in the steady-state configurations: $({\delta}^{z}_{+},{\delta}^{z}_{-}) = (0.1, 0.2)\gamma$ (the orientation not far from the lattice plane; narrow resonances), $(0.45, 1.75)\gamma$ (the orientation approximately normal to the plane; broad resonances).
The occupations of the eigenmodes for the two dipolar orientations of the (b) spectra are shown in (c) for the collective subradiant eigenmode where the atomic dipoles are coherently oscillating in the $x$ direction
(dashed-dotted lines), the collective eigenmode where the atomic dipoles are coherently oscillating in the $y$ direction (solid lines), the sum of the mode populations of all the other 1198 eigenmodes (dashed lines).  
The three mode population curves in (c) (green and cyan) for the broad resonance of (b) have the same location for the resonance and the same broad shape as the resonance in (b). The three narrow resonance population curves in (c) (red and blue) similarly follow precisely the narrow resonance in (b).		 
Both the narrow and the broad resonanced in (b) represent the large occupations of the subradiant mode and low occupations of the coherent in-plane mode, and the spectral shape closely follows the occupation profiles of (c).		 
The scattering resonance is approximately at the effective resonance of the subradiant mode [Eq.~\eqref{coll2modex}] $\Delta_P-\tilde\delta\simeq0$. }
		\label{fig:spectrum}
\end{figure*}

In Fig.~\ref{fig:spectrum}(a,b)
we show the spectra of the steady-state response of the forward or back scattered light into a narrow cone of $|\sin\theta|\alt 0.1$ for an incident plane wave. The full numerical simulation is again compared with the two-mode super-atom model of Eq.~\eqref{bothtwo}.
The super-atom model qualitatively captures the main features of the spectra, indicating that the resonance behavior is dominated by the two collective modes.
The spectra in large lattices exhibit Fano resonances due to a destructive interference between different scattering paths
that involve either the excitation  $ \Pc_I$ only, or a scattering via $ \Pc_P$, as in $ \Pc_I \rightarrow \Pc_P\rightarrow \Pc_I$,  and the interference is analogous to the interference of bright and dark modes in EIT.

The scattering of light in the exact forward and back directions is suppressed according to \EQREF{eq:transmission} when $\bar\delta^2\gg \upsilon_P\upsilon_I$. The resonances correspond to high (low) occupations of $\Pc_P$ ($\Pc_I$) excitations.
The ratio of the amplitudes in the steady-state solution of Eqs.~\eqref{eq:ssI} and~\eqref{eq:ssIb} indicates when the subradiant excitation $\Pc_P$ becomes dominant.
At the resonance $\Delta_P - \tilde{\delta} = 0$ we have $\Pc_P/\Pc_I=-\bar\delta/\upsilon_P$, and (assuming that $|\delta_P - \delta_I |\ll \upsilon_I$)
\begin{equation}
      \Pc_P = -\frac{\bar{\delta}}{\bar{\delta}^{2}+\upsilon_I\upsilon_P}f , \quad
		\Pc_I = \frac{\upsilon_P}{\bar{\delta}^{2}+\upsilon_I\upsilon_P}f \,.
\end{equation}
In Fig.~\ref{fig:spectrum}(c) we display the measure of the eigenmode populations [obtained using  \EQREF{eq:measure}] of the main eigenmodes in the steady-state response as a function of the incident light frequency. These populations correspond to the scattered light spectra shown in Fig.~\ref{fig:spectrum}(b). The occupations profiles are closely linked to the spectral profiles. The peak of the subradiant mode excitation (and the trough in the coherent in-plane mode excitation) represents the case where the light scattering in the forward or back direction vanishes, and the lattice becomes transparent. As explained earlier, in the limit of an infinitely large lattice with a subwavelength lattice spacing and fixed atomic positions only the exact forward or back scattering is possible, since in that case only the zeroth order Bragg peak survives.

The resonances strongly depend on the lattice size.
For the 3$\times$ 3 lattice [Fig.~\ref{fig:spectrum}(a)] the reflectance is only suppressed when $({\delta}^{z}_{+},{\delta}^{z}_{-}) = (0.45, 1.75)\gamma$, but not in the case of $({\delta}^{z}_{+},{\delta}^{z}_{-}) = (0.1, 0.2)\gamma$ when $\bar{\delta}^{2} \sim \upsilon_P\upsilon_I$ and the condition of \EQREF{eq:transmission} is not valid.

We can express the power reflectance $|r|^2$ in the limit $\upsilon_P/\bar{\delta} \simeq 0$
\begin{equation}
  \begin{split}
  \label{eq:no}
 |r|^2 \simeq  \frac{{(r_0\upsilon_I)}^{2}\[(\Delta_{P}-\tilde{\delta})^{2}+\upsilon_P^2\]}{|(\Delta_{P}-\tilde{\delta})^{4}-(\upsilon_I^2-2\bar{\delta}^{2})(\Delta_{P}-\tilde{\delta})^{2}+ \bar{\delta}^{4}|} \textrm{ ,}
  \end{split}
\end{equation}
where we have assumed $|\delta_P - \delta_I |\ll \upsilon_I$. In this limit we can then analytically calculate the half-width at half maximum for this resonance and obtain
\begin{equation}
	w \simeq \frac{1}{2}\(\sqrt{\upsilon_I^2 + 4\bar{\delta}^{2}} - \upsilon_I\)\,.
	\label{width}
\end{equation}
This simple expression qualitatively explains the observed behavior of the spectral resonances. If $\Pc_P$ is strongly excited by the Zeeman shifts $\bar{\delta}$, the resonance notably broadens. In the limit of a weak driving, the resonance narrows and eventually only depends on the very narrow resonance linewidth $\upsilon_P$, being a direct
manifestation of subradiance.

\subsubsection{Pulse delay}

We can also use the super-atom mode to calculate the group delay $\tau_g$ of a resonant light pulse for the 2D atom array due to cooperative response. This is given by the derivative of the
phase of the transmission amplitude with respect to the frequency
\beq
\tau_g = \left.\frac{d}{d\omega} \arg
    t(\omega)\right|_{\Delta_P-\tilde\delta=0}  \simeq {\upsilon_I\over \bar\delta^2}\,,
\eeq
where we have taken the limit of a large array and assumed $\bar\delta^2 \gg \upsilon_P \upsilon_I$ with $\delta_P \simeq \delta_I$. 
In the limit of a large lattice $\upsilon_P\rightarrow0$, and is generally very small, indicating the condition $\bar\delta^2 \gg \upsilon_P \upsilon_I$ is not particularly restrictive and  we can still simultaneously have a very small $\bar\delta \ll \upsilon_I$. This corresponds to an extremely narrow transmission resonance in which case the sample is transparent at the exact resonance with a sharp variation of the dispersion curve.

By using the value of $\upsilon_P\sim 10^{-3}\gamma$ for the $35\times35$  lattice~\cite{Facchinetti}, we can obtain a sharp resonance even for $\bar\delta^2\sim 10^{-2} \gamma^2$, and $\tau_g\sim 100/\gamma$ (assuming that $\upsilon_I\sim \gamma$). The cooperative effect is even more dramatic for the $150\times 150$ array for which $\upsilon_P\sim 10^{-4}\gamma$ (obtained from $\upsilon_P/\gamma\sim N^{-0.91}$~\cite{Facchinetti}), and
the group delay becomes significant $\tau_g\sim 10^3/\gamma$ even for an ultrathin 2D plane of atoms.

\subsubsection{The effects of atom confinement}

The atomic positions fluctuate due to the finite size of the trap. These can be taken into account in the simulations, as explained in Ref.~\cite{Jenkins2012a}.
For instance, in an optical lattice potential in the lowest vibrational level of each lattice site, the root-mean-square width of the wave function $\ell=\ell_y=\ell_z= as^{-1/4} / \pi$,
where $E_R = \pi^2\hbar^2/(2ma^2)$ is the lattice-photon recoil energy and each site has a potential depth
$sE_R$~\cite{Morsch06}. The current experiments in optical lattices with the Mott-insulator state can reach the depth $s\simeq 3000$~\cite{com:gross}.
The fluctuations of the atomic positions suppress the linewidth narrowing and, e.g., $s=50$ lattice for the example considered in Sec.~\ref{sec:comparison}
has the large array limit $\upsilon_P \simeq 0.15\gamma$. Using tight confinement  in the Lamb-Dicke regime $\ell_j\ll a$, e.g., by optical tweezers by means of nanofabrication, could further increase the lifetime of the subradiant state in large systems.

\section{Cooperative magnetometry}
\label{sec:magneto}

We have demonstrated how the collective eigenmodes of the atomic dipoles in the planar lattice exhibit behavior that is reminiscent of independent atom EIT.
In this section we show how the excitation of the collective subradiant state and the resulting narrow transmission resonance could be used in the detection of weak magnetic fields. The basic idea is closely related to the EIT magnetometry, based on the sharp dispersion in a transparent media~\cite{Fleisch_magneto,FleischhauerEtAlRMP2005}.
However, as in the EIT magnetometry the atoms are weakly interacting and can be described by the independent atom scattering model, the characteristic frequency for the suppression of absorption and the sharp dispersion is determined by the \emph{single-atom} resonance linewidth. The key element in the cooperative transmission is the interplay between the \emph{collective} subradiant excitation $ \Pc_P$ and strongly radiating $ \Pc_I$. In the large arrays, the corresponding resonance linewidths satisfy $\upsilon_P \ll \upsilon_I$. This can lead to very narrow transmission resonances, determined by the characteristic frequency $\upsilon_P$.

\subsection{Independent atom EIT magnetometry}

We first describe the standard independent-atom EIT magnetometry~\cite{Fleisch_magneto,FleischhauerEtAlRMP2005}.
We consider  a lambda-three-level EIT system, shown in Fig.~\ref{figure_eit}.
The destructive interference between the different transition amplitudes leads to a spectral transparency window where the absorption is suppressed. This is a single-atom effect where the atoms are assumed to respond to light independently.  For a small two-photon detuning between the levels $|a\>$ and $|b\>$, the electric susceptibility of the medium can be approximated by 
\beq
\chi=\chi'+i \chi''\,,
\eeq
 with
\beq
\chi'  \simeq -{24\pi\gamma_{bc}\rho\Delta\omega_{ab}\over k^3\Omega_c^2},\quad
\chi'' \simeq {12\pi \gamma_{bc}\rho \Gamma_{ab}\over k^3\Omega_c^2}\,,
\eeq
where the imaginary part $\chi''$ leads to absorption and the real part $\chi'$ to dispersion. Here $\rho$ denotes the density of the atoms, $\Delta\omega_{ab}$ the two-photon detuning between the levels $|a\>$ and $|b\>$, and $\gamma_{bc}$ the single-atom resonance linewidth of the probe $|b\>\leftrightarrow |c\>$ transition. The decay rate for the loss of coherence between the levels $|a\>$ and $|b\>$ is given by  $\Gamma_{ab}$ and it results, e.g., from atomic collisions and the fluctuations of the laser.
Since the two-photon detuning depends on the magnetic Zeeman shifts between the levels $\Delta\omega_{ab}\simeq \mu_B(m_a g_a-m_b g_b)/\hbar$ (here $\mu_B$ is Bohr's magneton, $g_j$ is the Land\'{e} g-factor and $m_j$ is the magnetic quantum number for the level $j$), the dispersion is proportional to the applied magnetic field, while the absorption at the same time is suppressed by EIT.
The phase shift of the light propagating a distance $L$ inside the sample is obtained by
\beq
\Delta \phi= k(n-1)L\simeq  -{2 \gamma_{bc}\Delta\omega_{ab}\over \Omega_c^2} \rho\sigma_{\rm cr} L\,,
\eeq
where we have used $n\simeq 1+\chi'/2$ for the index of refraction $n$, and $\sigma_{\rm cr}= 3\lambda^2/(2\pi)$ denotes the resonance cross-section.

The EIT magnetometry is based on achieving a large phase shift by propagating the light beam through a sufficiently large atom cloud, even when the coupling field intensity, proportional to $\Omega_c^2$, is large. Close to the EIT resonance an otherwise opaque medium becomes almost transparent due to the destructive interference between the different atomic transitions. This results in the suppressed attenuation of the beam (given by $\chi''$) and allows a long propagation distance inside the sample.
The suppression of the absorption at the EIT resonance is associated with a sharp variation of the dispersion curve.

\subsection{Super-atom model for cooperative magnetometry}

We propose a model for a cooperative magnetometry using stacked layers of 2D planar lattices of atoms. Within each lattice the atoms strongly interact via the scattered light (with subwavelength lattice spacing).
Here we assume that the different layers are not very close to each other and are separated by a distance more than the wavelength of the resonant light $\lambda$. Instead of fully treating the entire 3D lattice~\cite{Castin13}, we can then
as a first approximation ignore the recurrent scattering between the layers and add the interacting layers together analogously to the independent scattering approximation~\cite{Ruostekoski_waveguide}. In the model each planar lattice is then treated as a single super-atom and the interaction between the super-atoms (different lattices) is considered weak, such that super-atoms are modelled within the independent scattering approximation. 

In the limit of a large lattice and with not very closely-spaced planar lattices the $ \Pc_P$ excitation does not contribute to the coherent light propagation (since only the zeroth order Bragg peak exists in the radiation zone), and we write the effective electric susceptibility as 
\beq
\chi \simeq {\< P\> \over \epsilon_0 {\cal E}_0} \simeq {\rho {\cal D} \<{\cal P}_I\>\over \epsilon_0 {\cal E}_0}\,, 
\eeq
where the atom density can be given in terms of the layer separation $\Delta z$ as $\rho\simeq 1/(a^2\Delta z)$.
We assume that $\bar\delta^2 \gg \upsilon_P \upsilon_I$ and that the effective detuning from the $ \Pc_P$ excitation resonance $|\Delta_P-\tilde\delta|\ll \bar\delta$. Note, however, that an extremely narrow transmission resonance (at which the sample is transparent and exhibits a sharp variation of the dispersion curve) can be achieved
since in the limit of a large array, $\upsilon_P\rightarrow0$, and we can simultaneously have a very small $\bar\delta \ll \upsilon_I,\gamma$. 

We ignore the differences between the resonance frequencies of the two modes $\delta_P \simeq \delta_I$. From the two-mode model solutions we then find
\beq
\chi'  \simeq {6\pi\gamma\rho(\Delta_P-\tilde\delta)\over k^3\bar \delta^2},\quad
\chi'' \simeq {6\pi\gamma\rho \upsilon_P\over k^3\bar \delta^2}\,.
\eeq
The corresponding phase shift is given by
\beq
\Delta \phi =  {\gamma (\Delta_P-\tilde\delta)\over \bar \delta^2} \rho \sigma_{\rm cr} L\,.
\eeq
Here any changes in both $\bar\delta$ and $\tilde\delta$ are sensitive to the magnetic field variation. We can set the maximum allowed length of the sample to be defined by limiting the attenuation of the beam by absorption [given by $\exp(-\pi\chi'' L /\pi)$]  to be
$\pi\chi'' L_{\rm max}  /\pi \sim 1$. For $L_{\rm max}$ the phase shift $\Delta\phi\sim (\Delta_P-\tilde\delta)/\upsilon_P$ can still be very large due to extremely narrow subradiant mode resonance $\upsilon_P$.

\section{Concluding remarks}
\label{sec:conc}

Planar arrays of dipolar point emitters can respond strongly to light, exhibiting collective behavior. We have shown that this response can be qualitative captured by a simple model of a super-atom that shares features similar to those of independent-atom EIT. The super-atom model provides an intuitive, semi-analytic  description for the optical response of the lattice, including those of the complete reflection, full transmission, and the sharp Fano resonances in the spectral response, which were obtained in Refs.~\cite{CAIT,Facchinetti}.

The analysis of the lattice responses paves a way for harnessing the strong dipole-dipole interactions between the atoms. The collective  light-mediated interactions can be utilized in the
controlled preparation of a giant subradiant excitation in a single eigenmode where the correlated many-atom excitation spatially extends over the entire lattice and where the interactions strongly suppress radiative losses. Although super-radiance has been extensively studied in experiments~\cite{GrossHarochePhysRep1982} -- with the recent experiments focusing
on light in confined geometries~\cite{kimblesuper,Solano_super}, weak excitation regime~\cite{Roof16,Araujo16,wilkowski}, and the related shifts of the resonance frequencies~\cite{Keaveney2012,Meir13,ROH10,Jenkins_thermshift,Jennewein_trans,Dalibard_slab} -- subradiance generally has been elusive, due to its weak coupling to light.

Moreover, we proposed a scheme for a cooperative magnetometry.
Unlike in a magnetometry using EIT~\cite{Fleisch_magneto} in weakly interacting vapors that can be described by independent scattering approximation, the sensitivity to weak magnetic fields here is based on the strong light-induced dipole-dipole interactions between the atoms. As a result,  the width of the resonance is not limited by the single atom resonance linewidth, but by the much narrower collective subradiant linewidth.

\acknowledgments
We acknowledge financial support from  EPSRC.


\begin{thebibliography}{68}%
\makeatletter
\providecommand \@ifxundefined [1]{%
 \@ifx{#1\undefined}
}%
\providecommand \@ifnum [1]{%
 \ifnum #1\expandafter \@firstoftwo
 \else \expandafter \@secondoftwo
 \fi
}%
\providecommand \@ifx [1]{%
 \ifx #1\expandafter \@firstoftwo
 \else \expandafter \@secondoftwo
 \fi
}%
\providecommand \natexlab [1]{#1}%
\providecommand \enquote  [1]{``#1''}%
\providecommand \bibnamefont  [1]{#1}%
\providecommand \bibfnamefont [1]{#1}%
\providecommand \citenamefont [1]{#1}%
\providecommand \href@noop [0]{\@secondoftwo}%
\providecommand \href [0]{\begingroup \@sanitize@url \@href}%
\providecommand \@href[1]{\@@startlink{#1}\@@href}%
\providecommand \@@href[1]{\endgroup#1\@@endlink}%
\providecommand \@sanitize@url [0]{\catcode `\\12\catcode `\$12\catcode
  `\&12\catcode `\#12\catcode `\^12\catcode `\_12\catcode `\%12\relax}%
\providecommand \@@startlink[1]{}%
\providecommand \@@endlink[0]{}%
\providecommand \url  [0]{\begingroup\@sanitize@url \@url }%
\providecommand \@url [1]{\endgroup\@href {#1}{\urlprefix }}%
\providecommand \urlprefix  [0]{URL }%
\providecommand \Eprint [0]{\href }%
\providecommand \doibase [0]{http://dx.doi.org/}%
\providecommand \selectlanguage [0]{\@gobble}%
\providecommand \bibinfo  [0]{\@secondoftwo}%
\providecommand \bibfield  [0]{\@secondoftwo}%
\providecommand \translation [1]{[#1]}%
\providecommand \BibitemOpen [0]{}%
\providecommand \bibitemStop [0]{}%
\providecommand \bibitemNoStop [0]{.\EOS\space}%
\providecommand \EOS [0]{\spacefactor3000\relax}%
\providecommand \BibitemShut  [1]{\csname bibitem#1\endcsname}%
\let\auto@bib@innerbib\@empty
\bibitem [{\citenamefont {Fedotov}\ \emph {et~al.}(2010)\citenamefont
  {Fedotov}, \citenamefont {Papasimakis}, \citenamefont {Plum}, \citenamefont
  {Bitzer}, \citenamefont {Walther}, \citenamefont {Kuo}, \citenamefont
  {Tsai},\ and\ \citenamefont {Zheludev}}]{FedotovEtAlPRL2010}%
  \BibitemOpen
  \bibfield  {author} {\bibinfo {author} {\bibfnamefont {V.~A.}\ \bibnamefont
  {Fedotov}}, \bibinfo {author} {\bibfnamefont {N.}~\bibnamefont
  {Papasimakis}}, \bibinfo {author} {\bibfnamefont {E.}~\bibnamefont {Plum}},
  \bibinfo {author} {\bibfnamefont {A.}~\bibnamefont {Bitzer}}, \bibinfo
  {author} {\bibfnamefont {M.}~\bibnamefont {Walther}}, \bibinfo {author}
  {\bibfnamefont {P.}~\bibnamefont {Kuo}}, \bibinfo {author} {\bibfnamefont
  {D.~P.}\ \bibnamefont {Tsai}}, \ and\ \bibinfo {author} {\bibfnamefont
  {N.~I.}\ \bibnamefont {Zheludev}},\ }\bibfield  {title} {\enquote {\bibinfo
  {title} {Spectral collapse in ensembles of metamolecules},}\ }\href@noop {}
  {\bibfield  {journal} {\bibinfo  {journal} {Phys. Rev. Lett.}\ }\textbf
  {\bibinfo {volume} {104}},\ \bibinfo {pages} {223901} (\bibinfo {year}
  {2010})}\BibitemShut {NoStop}%
\bibitem [{\citenamefont {Lemoult}\ \emph {et~al.}(2010)\citenamefont
  {Lemoult}, \citenamefont {Lerosey}, \citenamefont {{de Rosny}},\ and\
  \citenamefont {Fink}}]{LemoultPRL10}%
  \BibitemOpen
  \bibfield  {author} {\bibinfo {author} {\bibfnamefont {Fabrice}\ \bibnamefont
  {Lemoult}}, \bibinfo {author} {\bibfnamefont {Geoffroy}\ \bibnamefont
  {Lerosey}}, \bibinfo {author} {\bibfnamefont {Julien}\ \bibnamefont {{de
  Rosny}}}, \ and\ \bibinfo {author} {\bibfnamefont {Mathias}\ \bibnamefont
  {Fink}},\ }\bibfield  {title} {\enquote {\bibinfo {title} {Resonant
  metalenses for breaking the diffraction barrier},}\ }\href {\doibase
  10.1103/PhysRevLett.104.203901} {\bibfield  {journal} {\bibinfo  {journal}
  {Phys. Rev. Lett.}\ }\textbf {\bibinfo {volume} {104}},\ \bibinfo {pages}
  {203901} (\bibinfo {year} {2010})}\BibitemShut {NoStop}%
\bibitem [{\citenamefont {Papasimakis}\ \emph {et~al.}(2009)\citenamefont
  {Papasimakis}, \citenamefont {Fu}, \citenamefont {Fedotov}, \citenamefont
  {Prosvirnin}, \citenamefont {Tsai},\ and\ \citenamefont
  {Zheludev}}]{PapasimakisEtAlAPL2009}%
  \BibitemOpen
  \bibfield  {author} {\bibinfo {author} {\bibfnamefont {N.}~\bibnamefont
  {Papasimakis}}, \bibinfo {author} {\bibfnamefont {Y.~H.}\ \bibnamefont {Fu}},
  \bibinfo {author} {\bibfnamefont {V.~A.}\ \bibnamefont {Fedotov}}, \bibinfo
  {author} {\bibfnamefont {S.~L.}\ \bibnamefont {Prosvirnin}}, \bibinfo
  {author} {\bibfnamefont {D.~P.}\ \bibnamefont {Tsai}}, \ and\ \bibinfo
  {author} {\bibfnamefont {N.~I.}\ \bibnamefont {Zheludev}},\ }\bibfield
  {title} {\enquote {\bibinfo {title} {Metamaterial with polarization and
  direction insensitive resonant transmission response mimicking
  electromagnetically induced transparency},}\ }\href@noop {} {\bibfield
  {journal} {\bibinfo  {journal} {Appl. Phys. Lett.}\ }\textbf {\bibinfo
  {volume} {94}},\ \bibinfo {pages} {211902} (\bibinfo {year}
  {2009})}\BibitemShut {NoStop}%
\bibitem [{\citenamefont {Jenkins}\ \emph {et~al.}(2017)\citenamefont
  {Jenkins}, \citenamefont {Ruostekoski}, \citenamefont {Papasimakis},
  \citenamefont {Savo},\ and\ \citenamefont {Zheludev}}]{Jenkins17}%
  \BibitemOpen
  \bibfield  {author} {\bibinfo {author} {\bibfnamefont {Stewart~D.}\
  \bibnamefont {Jenkins}}, \bibinfo {author} {\bibfnamefont {Janne}\
  \bibnamefont {Ruostekoski}}, \bibinfo {author} {\bibfnamefont {Nikitas}\
  \bibnamefont {Papasimakis}}, \bibinfo {author} {\bibfnamefont {Salvatore}\
  \bibnamefont {Savo}}, \ and\ \bibinfo {author} {\bibfnamefont {Nikolay~I.}\
  \bibnamefont {Zheludev}},\ }\bibfield  {title} {\enquote {\bibinfo {title}
  {Many-body subradiant excitations in metamaterial arrays: Experiment and
  theory},}\ }\href {\doibase 10.1103/PhysRevLett.119.053901} {\bibfield
  {journal} {\bibinfo  {journal} {Phys. Rev. Lett.}\ }\textbf {\bibinfo
  {volume} {119}},\ \bibinfo {pages} {053901} (\bibinfo {year}
  {2017})}\BibitemShut {NoStop}%
\bibitem [{\citenamefont {Jenkins}\ and\ \citenamefont
  {Ruostekoski}(2012{\natexlab{a}})}]{JenkinsLineWidthNJP}%
  \BibitemOpen
  \bibfield  {author} {\bibinfo {author} {\bibfnamefont {S.~D.}\ \bibnamefont
  {Jenkins}}\ and\ \bibinfo {author} {\bibfnamefont {J.}~\bibnamefont
  {Ruostekoski}},\ }\bibfield  {title} {\enquote {\bibinfo {title} {Cooperative
  resonance linewidth narrowing in a planar metamaterial},}\ }\href@noop {}
  {\bibfield  {journal} {\bibinfo  {journal} {New Journal of Physics}\ }\textbf
  {\bibinfo {volume} {14}},\ \bibinfo {pages} {103003} (\bibinfo {year}
  {2012}{\natexlab{a}})}\BibitemShut {NoStop}%
\bibitem [{\citenamefont {Jenkins}\ and\ \citenamefont
  {Ruostekoski}(2013)}]{CAIT}%
  \BibitemOpen
  \bibfield  {author} {\bibinfo {author} {\bibfnamefont {S.~D.}\ \bibnamefont
  {Jenkins}}\ and\ \bibinfo {author} {\bibfnamefont {J.}~\bibnamefont
  {Ruostekoski}},\ }\bibfield  {title} {\enquote {\bibinfo {title}
  {Metamaterial transparency induced by cooperative electromagnetic
  interactions},}\ }\href@noop {} {\bibfield  {journal} {\bibinfo  {journal}
  {Phys. Rev. Lett.}\ }\textbf {\bibinfo {volume} {111}},\ \bibinfo {pages}
  {147401} (\bibinfo {year} {2013})}\BibitemShut {NoStop}%
\bibitem [{\citenamefont {Weitenberg}\ \emph {et~al.}(2011)\citenamefont
  {Weitenberg}, \citenamefont {Endres}, \citenamefont {Sherson}, \citenamefont
  {Cheneau}, \citenamefont {Schau\ss}, \citenamefont {Fukuhara}, \citenamefont
  {Bloch},\ and\ \citenamefont {Kuhr}}]{singlespin}%
  \BibitemOpen
  \bibfield  {author} {\bibinfo {author} {\bibfnamefont {Christof}\
  \bibnamefont {Weitenberg}}, \bibinfo {author} {\bibfnamefont {Manuel}\
  \bibnamefont {Endres}}, \bibinfo {author} {\bibfnamefont {Jacob~F.}\
  \bibnamefont {Sherson}}, \bibinfo {author} {\bibfnamefont {Marc}\
  \bibnamefont {Cheneau}}, \bibinfo {author} {\bibfnamefont {Peter}\
  \bibnamefont {Schau\ss}}, \bibinfo {author} {\bibfnamefont {Takeshi}\
  \bibnamefont {Fukuhara}}, \bibinfo {author} {\bibfnamefont {Immanuel}\
  \bibnamefont {Bloch}}, \ and\ \bibinfo {author} {\bibfnamefont {Stefan}\
  \bibnamefont {Kuhr}},\ }\bibfield  {title} {\enquote {\bibinfo {title}
  {Single-spin addressing in an atomic {Mott} insulator},}\ }\href@noop {}
  {\bibfield  {journal} {\bibinfo  {journal} {Nature}\ }\textbf {\bibinfo
  {volume} {471}},\ \bibinfo {pages} {319} (\bibinfo {year}
  {2011})}\BibitemShut {NoStop}%
\bibitem [{\citenamefont {Nogrette}\ \emph {et~al.}(2014)\citenamefont
  {Nogrette}, \citenamefont {Labuhn}, \citenamefont {Ravets}, \citenamefont
  {Barredo}, \citenamefont {B\'eguin}, \citenamefont {Vernier}, \citenamefont
  {Lahaye},\ and\ \citenamefont {Browaeys}}]{Browaeys_trap}%
  \BibitemOpen
  \bibfield  {author} {\bibinfo {author} {\bibfnamefont {F.}~\bibnamefont
  {Nogrette}}, \bibinfo {author} {\bibfnamefont {H.}~\bibnamefont {Labuhn}},
  \bibinfo {author} {\bibfnamefont {S.}~\bibnamefont {Ravets}}, \bibinfo
  {author} {\bibfnamefont {D.}~\bibnamefont {Barredo}}, \bibinfo {author}
  {\bibfnamefont {L.}~\bibnamefont {B\'eguin}}, \bibinfo {author}
  {\bibfnamefont {A.}~\bibnamefont {Vernier}}, \bibinfo {author} {\bibfnamefont
  {T.}~\bibnamefont {Lahaye}}, \ and\ \bibinfo {author} {\bibfnamefont
  {A.}~\bibnamefont {Browaeys}},\ }\bibfield  {title} {\enquote {\bibinfo
  {title} {Single-atom trapping in holographic 2d arrays of microtraps with
  arbitrary geometries},}\ }\href {\doibase 10.1103/PhysRevX.4.021034}
  {\bibfield  {journal} {\bibinfo  {journal} {Phys. Rev. X}\ }\textbf {\bibinfo
  {volume} {4}},\ \bibinfo {pages} {021034} (\bibinfo {year}
  {2014})}\BibitemShut {NoStop}%
\bibitem [{\citenamefont {Dicke}(1954)}]{Dicke54}%
  \BibitemOpen
  \bibfield  {author} {\bibinfo {author} {\bibfnamefont {R.~H.}\ \bibnamefont
  {Dicke}},\ }\bibfield  {title} {\enquote {\bibinfo {title} {Coherence in
  spontaneous radiation processes},}\ }\href {\doibase 10.1103/PhysRev.93.99}
  {\bibfield  {journal} {\bibinfo  {journal} {Phys. Rev.}\ }\textbf {\bibinfo
  {volume} {93}},\ \bibinfo {pages} {99--110} (\bibinfo {year}
  {1954})}\BibitemShut {NoStop}%
\bibitem [{\citenamefont {Jenkins}\ and\ \citenamefont
  {Ruostekoski}(2012{\natexlab{b}})}]{Jenkins2012a}%
  \BibitemOpen
  \bibfield  {author} {\bibinfo {author} {\bibfnamefont {Stewart~D.}\
  \bibnamefont {Jenkins}}\ and\ \bibinfo {author} {\bibfnamefont {Janne}\
  \bibnamefont {Ruostekoski}},\ }\bibfield  {title} {\enquote {\bibinfo {title}
  {Controlled manipulation of light by cooperative response of atoms in an
  optical lattice},}\ }\href {\doibase 10.1103/PhysRevA.86.031602} {\bibfield
  {journal} {\bibinfo  {journal} {Phys. Rev. A}\ }\textbf {\bibinfo {volume}
  {86}},\ \bibinfo {pages} {031602} (\bibinfo {year}
  {2012}{\natexlab{b}})}\BibitemShut {NoStop}%
\bibitem [{\citenamefont {Hebenstreit}\ \emph {et~al.}(2017)\citenamefont
  {Hebenstreit}, \citenamefont {Kraus}, \citenamefont {Ostermann},\ and\
  \citenamefont {Ritsch}}]{Ritsch_subr}%
  \BibitemOpen
  \bibfield  {author} {\bibinfo {author} {\bibfnamefont {Martin}\ \bibnamefont
  {Hebenstreit}}, \bibinfo {author} {\bibfnamefont {Barbara}\ \bibnamefont
  {Kraus}}, \bibinfo {author} {\bibfnamefont {Laurin}\ \bibnamefont
  {Ostermann}}, \ and\ \bibinfo {author} {\bibfnamefont {Helmut}\ \bibnamefont
  {Ritsch}},\ }\bibfield  {title} {\enquote {\bibinfo {title} {Subradiance via
  entanglement in atoms with several independent decay channels},}\ }\href
  {\doibase 10.1103/PhysRevLett.118.143602} {\bibfield  {journal} {\bibinfo
  {journal} {Phys. Rev. Lett.}\ }\textbf {\bibinfo {volume} {118}},\ \bibinfo
  {pages} {143602} (\bibinfo {year} {2017})}\BibitemShut {NoStop}%
\bibitem [{\citenamefont {Facchinetti}\ \emph {et~al.}(2016)\citenamefont
  {Facchinetti}, \citenamefont {Jenkins},\ and\ \citenamefont
  {Ruostekoski}}]{Facchinetti}%
  \BibitemOpen
  \bibfield  {author} {\bibinfo {author} {\bibfnamefont {G.}~\bibnamefont
  {Facchinetti}}, \bibinfo {author} {\bibfnamefont {S.~D.}\ \bibnamefont
  {Jenkins}}, \ and\ \bibinfo {author} {\bibfnamefont {J.}~\bibnamefont
  {Ruostekoski}},\ }\bibfield  {title} {\enquote {\bibinfo {title} {Storing
  light with subradiant correlations in arrays of atoms},}\ }\href {\doibase
  10.1103/PhysRevLett.117.243601} {\bibfield  {journal} {\bibinfo  {journal}
  {Phys. Rev. Lett.}\ }\textbf {\bibinfo {volume} {117}},\ \bibinfo {pages}
  {243601} (\bibinfo {year} {2016})}\BibitemShut {NoStop}%
\bibitem [{\citenamefont {Perczel}\ \emph {et~al.}(2017)\citenamefont
  {Perczel}, \citenamefont {Borregaard}, \citenamefont {Chang}, \citenamefont
  {Pichler}, \citenamefont {Yelin}, \citenamefont {Zoller},\ and\ \citenamefont
  {Lukin}}]{Perczel}%
  \BibitemOpen
  \bibfield  {author} {\bibinfo {author} {\bibfnamefont {J.}~\bibnamefont
  {Perczel}}, \bibinfo {author} {\bibfnamefont {J.}~\bibnamefont {Borregaard}},
  \bibinfo {author} {\bibfnamefont {D.~E.}\ \bibnamefont {Chang}}, \bibinfo
  {author} {\bibfnamefont {H.}~\bibnamefont {Pichler}}, \bibinfo {author}
  {\bibfnamefont {S.~F.}\ \bibnamefont {Yelin}}, \bibinfo {author}
  {\bibfnamefont {P.}~\bibnamefont {Zoller}}, \ and\ \bibinfo {author}
  {\bibfnamefont {M.~D.}\ \bibnamefont {Lukin}},\ }\bibfield  {title} {\enquote
  {\bibinfo {title} {Photonic band structure of two-dimensional atomic
  lattices},}\ }\href {\doibase 10.1103/PhysRevA.96.063801} {\bibfield
  {journal} {\bibinfo  {journal} {Phys. Rev. A}\ }\textbf {\bibinfo {volume}
  {96}},\ \bibinfo {pages} {063801} (\bibinfo {year} {2017})}\BibitemShut
  {NoStop}%
\bibitem [{\citenamefont {Bettles}\ \emph {et~al.}(2017)\citenamefont
  {Bettles}, \citenamefont {Min\'a\ifmmode~\check{r}\else \v{r}\fi{}},
  \citenamefont {Adams}, \citenamefont {Lesanovsky},\ and\ \citenamefont
  {Olmos}}]{Bettles_17topo}%
  \BibitemOpen
  \bibfield  {author} {\bibinfo {author} {\bibfnamefont {R.~J.}\ \bibnamefont
  {Bettles}}, \bibinfo {author} {\bibfnamefont {J.}~\bibnamefont
  {Min\'a\ifmmode~\check{r}\else \v{r}\fi{}}}, \bibinfo {author} {\bibfnamefont
  {C.~S.}\ \bibnamefont {Adams}}, \bibinfo {author} {\bibfnamefont
  {I.}~\bibnamefont {Lesanovsky}}, \ and\ \bibinfo {author} {\bibfnamefont
  {B.}~\bibnamefont {Olmos}},\ }\bibfield  {title} {\enquote {\bibinfo {title}
  {Topological properties of a dense atomic lattice gas},}\ }\href {\doibase
  10.1103/PhysRevA.96.041603} {\bibfield  {journal} {\bibinfo  {journal} {Phys.
  Rev. A}\ }\textbf {\bibinfo {volume} {96}},\ \bibinfo {pages} {041603}
  (\bibinfo {year} {2017})}\BibitemShut {NoStop}%
\bibitem [{\citenamefont {Bettles}\ \emph {et~al.}(2015)\citenamefont
  {Bettles}, \citenamefont {Gardiner},\ and\ \citenamefont
  {Adams}}]{Bettles_lattice}%
  \BibitemOpen
  \bibfield  {author} {\bibinfo {author} {\bibfnamefont {Robert~J.}\
  \bibnamefont {Bettles}}, \bibinfo {author} {\bibfnamefont {Simon~A.}\
  \bibnamefont {Gardiner}}, \ and\ \bibinfo {author} {\bibfnamefont
  {Charles~S.}\ \bibnamefont {Adams}},\ }\bibfield  {title} {\enquote {\bibinfo
  {title} {Cooperative ordering in lattices of interacting two-level
  dipoles},}\ }\href {\doibase 10.1103/PhysRevA.92.063822} {\bibfield
  {journal} {\bibinfo  {journal} {Phys. Rev. A}\ }\textbf {\bibinfo {volume}
  {92}},\ \bibinfo {pages} {063822} (\bibinfo {year} {2015})}\BibitemShut
  {NoStop}%
\bibitem [{\citenamefont {Yoo}\ and\ \citenamefont {Paik}(2016)}]{Yoo16}%
  \BibitemOpen
  \bibfield  {author} {\bibinfo {author} {\bibfnamefont {Sung-Mi}\ \bibnamefont
  {Yoo}}\ and\ \bibinfo {author} {\bibfnamefont {Sun~Mok}\ \bibnamefont
  {Paik}},\ }\bibfield  {title} {\enquote {\bibinfo {title} {Cooperative
  optical response of 2d dense lattices with strongly correlated dipoles},}\
  }\href {\doibase 10.1364/OE.24.002156} {\bibfield  {journal} {\bibinfo
  {journal} {Opt. Express}\ }\textbf {\bibinfo {volume} {24}},\ \bibinfo
  {pages} {2156--2165} (\bibinfo {year} {2016})}\BibitemShut {NoStop}%
\bibitem [{\citenamefont {Bettles}\ \emph {et~al.}(2016)\citenamefont
  {Bettles}, \citenamefont {Gardiner},\ and\ \citenamefont
  {Adams}}]{Bettles_prl16}%
  \BibitemOpen
  \bibfield  {author} {\bibinfo {author} {\bibfnamefont {Robert~J.}\
  \bibnamefont {Bettles}}, \bibinfo {author} {\bibfnamefont {Simon~A.}\
  \bibnamefont {Gardiner}}, \ and\ \bibinfo {author} {\bibfnamefont
  {Charles~S.}\ \bibnamefont {Adams}},\ }\bibfield  {title} {\enquote {\bibinfo
  {title} {Enhanced optical cross section via collective coupling of atomic
  dipoles in a 2d array},}\ }\href {\doibase 10.1103/PhysRevLett.116.103602}
  {\bibfield  {journal} {\bibinfo  {journal} {Phys. Rev. Lett.}\ }\textbf
  {\bibinfo {volume} {116}},\ \bibinfo {pages} {103602} (\bibinfo {year}
  {2016})}\BibitemShut {NoStop}%
\bibitem [{\citenamefont {Wang}\ \emph {et~al.}(2017)\citenamefont {Wang},
  \citenamefont {Zhao}, \citenamefont {Kan},\ and\ \citenamefont
  {Huang}}]{Wang17}%
  \BibitemOpen
  \bibfield  {author} {\bibinfo {author} {\bibfnamefont {B.~X.}\ \bibnamefont
  {Wang}}, \bibinfo {author} {\bibfnamefont {C.~Y.}\ \bibnamefont {Zhao}},
  \bibinfo {author} {\bibfnamefont {Y.~H.}\ \bibnamefont {Kan}}, \ and\
  \bibinfo {author} {\bibfnamefont {T.~C.}\ \bibnamefont {Huang}},\ }\bibfield
  {title} {\enquote {\bibinfo {title} {Design of metasurface polarizers based
  on two-dimensional cold atomic arrays},}\ }\href {\doibase
  10.1364/OE.25.018760} {\bibfield  {journal} {\bibinfo  {journal} {Opt.
  Express}\ }\textbf {\bibinfo {volume} {25}},\ \bibinfo {pages} {18760--18773}
  (\bibinfo {year} {2017})}\BibitemShut {NoStop}%
\bibitem [{\citenamefont {Plankensteiner}\ \emph {et~al.}(2017)\citenamefont
  {Plankensteiner}, \citenamefont {Sommer}, \citenamefont {Ritsch},\ and\
  \citenamefont {Genes}}]{Genes17}%
  \BibitemOpen
  \bibfield  {author} {\bibinfo {author} {\bibfnamefont {David}\ \bibnamefont
  {Plankensteiner}}, \bibinfo {author} {\bibfnamefont {Christian}\ \bibnamefont
  {Sommer}}, \bibinfo {author} {\bibfnamefont {Helmut}\ \bibnamefont {Ritsch}},
  \ and\ \bibinfo {author} {\bibfnamefont {Claudiu}\ \bibnamefont {Genes}},\
  }\bibfield  {title} {\enquote {\bibinfo {title} {Cavity antiresonance
  spectroscopy of dipole coupled subradiant arrays},}\ }\href {\doibase
  10.1103/PhysRevLett.119.093601} {\bibfield  {journal} {\bibinfo  {journal}
  {Phys. Rev. Lett.}\ }\textbf {\bibinfo {volume} {119}},\ \bibinfo {pages}
  {093601} (\bibinfo {year} {2017})}\BibitemShut {NoStop}%
\bibitem [{\citenamefont {Shahmoon}\ \emph {et~al.}(2017)\citenamefont
  {Shahmoon}, \citenamefont {Wild}, \citenamefont {Lukin},\ and\ \citenamefont
  {Yelin}}]{Shahmoon}%
  \BibitemOpen
  \bibfield  {author} {\bibinfo {author} {\bibfnamefont {Ephraim}\ \bibnamefont
  {Shahmoon}}, \bibinfo {author} {\bibfnamefont {Dominik~S.}\ \bibnamefont
  {Wild}}, \bibinfo {author} {\bibfnamefont {Mikhail~D.}\ \bibnamefont
  {Lukin}}, \ and\ \bibinfo {author} {\bibfnamefont {Susanne~F.}\ \bibnamefont
  {Yelin}},\ }\bibfield  {title} {\enquote {\bibinfo {title} {Cooperative
  resonances in light scattering from two-dimensional atomic arrays},}\ }\href
  {\doibase 10.1103/PhysRevLett.118.113601} {\bibfield  {journal} {\bibinfo
  {journal} {Phys. Rev. Lett.}\ }\textbf {\bibinfo {volume} {118}},\ \bibinfo
  {pages} {113601} (\bibinfo {year} {2017})}\BibitemShut {NoStop}%
\bibitem [{\citenamefont {Asenjo-Garcia}\ \emph {et~al.}(2017)\citenamefont
  {Asenjo-Garcia}, \citenamefont {Moreno-Cardoner}, \citenamefont {Albrecht},
  \citenamefont {Kimble},\ and\ \citenamefont {Chang}}]{Asenjo_prx}%
  \BibitemOpen
  \bibfield  {author} {\bibinfo {author} {\bibfnamefont {A.}~\bibnamefont
  {Asenjo-Garcia}}, \bibinfo {author} {\bibfnamefont {M.}~\bibnamefont
  {Moreno-Cardoner}}, \bibinfo {author} {\bibfnamefont {A.}~\bibnamefont
  {Albrecht}}, \bibinfo {author} {\bibfnamefont {H.~J.}\ \bibnamefont
  {Kimble}}, \ and\ \bibinfo {author} {\bibfnamefont {D.~E.}\ \bibnamefont
  {Chang}},\ }\bibfield  {title} {\enquote {\bibinfo {title} {Exponential
  improvement in photon storage fidelities using subradiance and ``selective
  radiance'' in atomic arrays},}\ }\href {\doibase 10.1103/PhysRevX.7.031024}
  {\bibfield  {journal} {\bibinfo  {journal} {Phys. Rev. X}\ }\textbf {\bibinfo
  {volume} {7}},\ \bibinfo {pages} {031024} (\bibinfo {year}
  {2017})}\BibitemShut {NoStop}%
\bibitem [{\citenamefont {Olmos}\ \emph {et~al.}(2013)\citenamefont {Olmos},
  \citenamefont {Yu}, \citenamefont {Singh}, \citenamefont {Schreck},
  \citenamefont {Bongs},\ and\ \citenamefont {Lesanovsky}}]{Olmos13}%
  \BibitemOpen
  \bibfield  {author} {\bibinfo {author} {\bibfnamefont {B.}~\bibnamefont
  {Olmos}}, \bibinfo {author} {\bibfnamefont {D.}~\bibnamefont {Yu}}, \bibinfo
  {author} {\bibfnamefont {Y.}~\bibnamefont {Singh}}, \bibinfo {author}
  {\bibfnamefont {F.}~\bibnamefont {Schreck}}, \bibinfo {author} {\bibfnamefont
  {K.}~\bibnamefont {Bongs}}, \ and\ \bibinfo {author} {\bibfnamefont
  {I.}~\bibnamefont {Lesanovsky}},\ }\bibfield  {title} {\enquote {\bibinfo
  {title} {Long-range interacting many-body systems with alkaline-earth-metal
  atoms},}\ }\href {\doibase 10.1103/PhysRevLett.110.143602} {\bibfield
  {journal} {\bibinfo  {journal} {Phys. Rev. Lett.}\ }\textbf {\bibinfo
  {volume} {110}},\ \bibinfo {pages} {143602} (\bibinfo {year}
  {2013})}\BibitemShut {NoStop}%
\bibitem [{\citenamefont {Fleischhauer}\ \emph {et~al.}(2005)\citenamefont
  {Fleischhauer}, \citenamefont {Imamoglu},\ and\ \citenamefont
  {Marangos}}]{FleischhauerEtAlRMP2005}%
  \BibitemOpen
  \bibfield  {author} {\bibinfo {author} {\bibfnamefont {M.}~\bibnamefont
  {Fleischhauer}}, \bibinfo {author} {\bibfnamefont {A.}~\bibnamefont
  {Imamoglu}}, \ and\ \bibinfo {author} {\bibfnamefont {J.~P.}\ \bibnamefont
  {Marangos}},\ }\bibfield  {title} {\enquote {\bibinfo {title}
  {Electromagnetically induced transparency: Optics in coherent media},}\
  }\href@noop {} {\bibfield  {journal} {\bibinfo  {journal} {Rev. Mod. Phys.}\
  }\textbf {\bibinfo {volume} {77}},\ \bibinfo {pages} {633--673} (\bibinfo
  {year} {2005})}\BibitemShut {NoStop}%
\bibitem [{\citenamefont {DeVoe}\ and\ \citenamefont {Brewer}(1996)}]{DeVoe}%
  \BibitemOpen
  \bibfield  {author} {\bibinfo {author} {\bibfnamefont {R.~G.}\ \bibnamefont
  {DeVoe}}\ and\ \bibinfo {author} {\bibfnamefont {R.~G.}\ \bibnamefont
  {Brewer}},\ }\bibfield  {title} {\enquote {\bibinfo {title} {Observation of
  superradiant and subradiant spontaneous emission of two trapped ions},}\
  }\href {\doibase 10.1103/PhysRevLett.76.2049} {\bibfield  {journal} {\bibinfo
   {journal} {Phys. Rev. Lett.}\ }\textbf {\bibinfo {volume} {76}},\ \bibinfo
  {pages} {2049--2052} (\bibinfo {year} {1996})}\BibitemShut {NoStop}%
\bibitem [{\citenamefont {Hettich}\ \emph {et~al.}(2002)\citenamefont
  {Hettich}, \citenamefont {Schmitt}, \citenamefont {Zitzmann}, \citenamefont
  {K�hn}, \citenamefont {Gerhardt},\ and\ \citenamefont
  {Sandoghdar}}]{Hettich}%
  \BibitemOpen
  \bibfield  {author} {\bibinfo {author} {\bibfnamefont {C.}~\bibnamefont
  {Hettich}}, \bibinfo {author} {\bibfnamefont {C.}~\bibnamefont {Schmitt}},
  \bibinfo {author} {\bibfnamefont {J.}~\bibnamefont {Zitzmann}}, \bibinfo
  {author} {\bibfnamefont {S.}~\bibnamefont {K�hn}}, \bibinfo {author}
  {\bibfnamefont {I.}~\bibnamefont {Gerhardt}}, \ and\ \bibinfo {author}
  {\bibfnamefont {V.}~\bibnamefont {Sandoghdar}},\ }\bibfield  {title}
  {\enquote {\bibinfo {title} {Nanometer resolution and coherent optical dipole
  coupling of two individual molecules},}\ }\href {\doibase
  10.1126/science.1075606} {\bibfield  {journal} {\bibinfo  {journal}
  {Science}\ }\textbf {\bibinfo {volume} {298}},\ \bibinfo {pages} {385--389}
  (\bibinfo {year} {2002})}\BibitemShut {NoStop}%
\bibitem [{\citenamefont {McGuyer}\ \emph {et~al.}(2015)\citenamefont
  {McGuyer}, \citenamefont {McDonald}, \citenamefont {Iwata}, \citenamefont
  {Tarallo}, \citenamefont {Skomorowski}, \citenamefont {Moszynski},\ and\
  \citenamefont {Zelevinsky}}]{McGuyer}%
  \BibitemOpen
  \bibfield  {author} {\bibinfo {author} {\bibfnamefont {B.~H.}\ \bibnamefont
  {McGuyer}}, \bibinfo {author} {\bibfnamefont {M.}~\bibnamefont {McDonald}},
  \bibinfo {author} {\bibfnamefont {G.~Z.}\ \bibnamefont {Iwata}}, \bibinfo
  {author} {\bibfnamefont {M.~G.}\ \bibnamefont {Tarallo}}, \bibinfo {author}
  {\bibfnamefont {W.}~\bibnamefont {Skomorowski}}, \bibinfo {author}
  {\bibfnamefont {R.}~\bibnamefont {Moszynski}}, \ and\ \bibinfo {author}
  {\bibfnamefont {T.}~\bibnamefont {Zelevinsky}},\ }\bibfield  {title}
  {\enquote {\bibinfo {title} {Precise study of asymptotic physics with
  subradiant ultracold molecules},}\ }\href {\doibase 10.1038/NPHYS3182}
  {\bibfield  {journal} {\bibinfo  {journal} {Nat. Phys.}\ }\textbf {\bibinfo
  {volume} {11}},\ \bibinfo {pages} {32--36} (\bibinfo {year}
  {2015})}\BibitemShut {NoStop}%
\bibitem [{\citenamefont {Takasu}\ \emph {et~al.}(2012)\citenamefont {Takasu},
  \citenamefont {Saito}, \citenamefont {Takahashi}, \citenamefont {Borkowski},
  \citenamefont {Ciury\l{}o},\ and\ \citenamefont {Julienne}}]{Takasu}%
  \BibitemOpen
  \bibfield  {author} {\bibinfo {author} {\bibfnamefont {Yosuke}\ \bibnamefont
  {Takasu}}, \bibinfo {author} {\bibfnamefont {Yutaka}\ \bibnamefont {Saito}},
  \bibinfo {author} {\bibfnamefont {Yoshiro}\ \bibnamefont {Takahashi}},
  \bibinfo {author} {\bibfnamefont {Mateusz}\ \bibnamefont {Borkowski}},
  \bibinfo {author} {\bibfnamefont {Roman}\ \bibnamefont {Ciury\l{}o}}, \ and\
  \bibinfo {author} {\bibfnamefont {Paul~S.}\ \bibnamefont {Julienne}},\
  }\bibfield  {title} {\enquote {\bibinfo {title} {Controlled production of
  subradiant states of a diatomic molecule in an optical lattice},}\ }\href
  {\doibase 10.1103/PhysRevLett.108.173002} {\bibfield  {journal} {\bibinfo
  {journal} {Phys. Rev. Lett.}\ }\textbf {\bibinfo {volume} {108}},\ \bibinfo
  {pages} {173002} (\bibinfo {year} {2012})}\BibitemShut {NoStop}%
\bibitem [{\citenamefont {Guerin}\ \emph {et~al.}(2016)\citenamefont {Guerin},
  \citenamefont {Ara\'ujo},\ and\ \citenamefont {Kaiser}}]{Guerin_subr16}%
  \BibitemOpen
  \bibfield  {author} {\bibinfo {author} {\bibfnamefont {William}\ \bibnamefont
  {Guerin}}, \bibinfo {author} {\bibfnamefont {Michelle~O.}\ \bibnamefont
  {Ara\'ujo}}, \ and\ \bibinfo {author} {\bibfnamefont {Robin}\ \bibnamefont
  {Kaiser}},\ }\bibfield  {title} {\enquote {\bibinfo {title} {Subradiance in a
  large cloud of cold atoms},}\ }\href {\doibase
  10.1103/PhysRevLett.116.083601} {\bibfield  {journal} {\bibinfo  {journal}
  {Phys. Rev. Lett.}\ }\textbf {\bibinfo {volume} {116}},\ \bibinfo {pages}
  {083601} (\bibinfo {year} {2016})}\BibitemShut {NoStop}%
\bibitem [{\citenamefont {Hammerer}\ \emph {et~al.}(2010)\citenamefont
  {Hammerer}, \citenamefont {S\o{}rensen},\ and\ \citenamefont
  {Polzik}}]{HAM10}%
  \BibitemOpen
  \bibfield  {author} {\bibinfo {author} {\bibfnamefont {Klemens}\ \bibnamefont
  {Hammerer}}, \bibinfo {author} {\bibfnamefont {Anders~S.}\ \bibnamefont
  {S\o{}rensen}}, \ and\ \bibinfo {author} {\bibfnamefont {Eugene~S.}\
  \bibnamefont {Polzik}},\ }\bibfield  {title} {\enquote {\bibinfo {title}
  {Quantum interface between light and atomic ensembles},}\ }\href {\doibase
  10.1103/RevModPhys.82.1041} {\bibfield  {journal} {\bibinfo  {journal} {Rev.
  Mod. Phys.}\ }\textbf {\bibinfo {volume} {82}},\ \bibinfo {pages}
  {1041--1093} (\bibinfo {year} {2010})}\BibitemShut {NoStop}%
\bibitem [{\citenamefont {Nicholson}\ \emph {et~al.}(2015)\citenamefont
  {Nicholson}, \citenamefont {Campbell}, \citenamefont {Hutson}, \citenamefont
  {Marti}, \citenamefont {Bloom}, \citenamefont {McNally}, \citenamefont
  {Zhang}, \citenamefont {Barrett}, \citenamefont {Safronova}, \citenamefont
  {Strouse}, \citenamefont {Tew},\ and\ \citenamefont {Ye}}]{Nicholson_clock}%
  \BibitemOpen
  \bibfield  {author} {\bibinfo {author} {\bibfnamefont {T.~L.}\ \bibnamefont
  {Nicholson}}, \bibinfo {author} {\bibfnamefont {S.~L.}\ \bibnamefont
  {Campbell}}, \bibinfo {author} {\bibfnamefont {R.~B.}\ \bibnamefont
  {Hutson}}, \bibinfo {author} {\bibfnamefont {G.~E.}\ \bibnamefont {Marti}},
  \bibinfo {author} {\bibfnamefont {B.~J.}\ \bibnamefont {Bloom}}, \bibinfo
  {author} {\bibfnamefont {R.~L.}\ \bibnamefont {McNally}}, \bibinfo {author}
  {\bibfnamefont {W.}~\bibnamefont {Zhang}}, \bibinfo {author} {\bibfnamefont
  {M.~D.}\ \bibnamefont {Barrett}}, \bibinfo {author} {\bibfnamefont {M.~S.}\
  \bibnamefont {Safronova}}, \bibinfo {author} {\bibfnamefont {G.~F.}\
  \bibnamefont {Strouse}}, \bibinfo {author} {\bibfnamefont {W.~L.}\
  \bibnamefont {Tew}}, \ and\ \bibinfo {author} {\bibfnamefont
  {J.}~\bibnamefont {Ye}},\ }\bibfield  {title} {\enquote {\bibinfo {title}
  {Systematic evaluation of an atomic clock at 2$\times10^{-18}$ total
  uncertainty},}\ }\href {http://dx.doi.org/10.1038/ncomms7896} {\bibfield
  {journal} {\bibinfo  {journal} {Nat Commun}\ }\textbf {\bibinfo {volume}
  {6}},\ \bibinfo {pages} {6896} (\bibinfo {year} {2015})}\BibitemShut
  {NoStop}%
\bibitem [{\citenamefont {Bromley}\ \emph {et~al.}(2016)\citenamefont
  {Bromley}, \citenamefont {Zhu}, \citenamefont {Bishof}, \citenamefont
  {Zhang}, \citenamefont {Bothwell}, \citenamefont {Schachenmayer},
  \citenamefont {Nicholson}, \citenamefont {Kaiser}, \citenamefont {Yelin},
  \citenamefont {Lukin}, \citenamefont {Rey},\ and\ \citenamefont
  {Ye}}]{Ye2016}%
  \BibitemOpen
  \bibfield  {author} {\bibinfo {author} {\bibfnamefont {S.~L.}\ \bibnamefont
  {Bromley}}, \bibinfo {author} {\bibfnamefont {B.}~\bibnamefont {Zhu}},
  \bibinfo {author} {\bibfnamefont {M.}~\bibnamefont {Bishof}}, \bibinfo
  {author} {\bibfnamefont {X.}~\bibnamefont {Zhang}}, \bibinfo {author}
  {\bibfnamefont {T.}~\bibnamefont {Bothwell}}, \bibinfo {author}
  {\bibfnamefont {J.}~\bibnamefont {Schachenmayer}}, \bibinfo {author}
  {\bibfnamefont {T.~L.}\ \bibnamefont {Nicholson}}, \bibinfo {author}
  {\bibfnamefont {R.}~\bibnamefont {Kaiser}}, \bibinfo {author} {\bibfnamefont
  {S.~F.}\ \bibnamefont {Yelin}}, \bibinfo {author} {\bibfnamefont {M.~D.}\
  \bibnamefont {Lukin}}, \bibinfo {author} {\bibfnamefont {A.~M.}\ \bibnamefont
  {Rey}}, \ and\ \bibinfo {author} {\bibfnamefont {J.}~\bibnamefont {Ye}},\
  }\bibfield  {title} {\enquote {\bibinfo {title} {Collective atomic scattering
  and motional effects in a dense coherent medium},}\ }\href
  {http://dx.doi.org/10.1038/ncomms11039} {\bibfield  {journal} {\bibinfo
  {journal} {Nat Commun}\ }\textbf {\bibinfo {volume} {7}},\ \bibinfo {pages}
  {11039} (\bibinfo {year} {2016})}\BibitemShut {NoStop}%
\bibitem [{\citenamefont {Budker}\ and\ \citenamefont
  {Romalis}(2007)}]{BudkerRamalisNatPhys2007}%
  \BibitemOpen
  \bibfield  {author} {\bibinfo {author} {\bibfnamefont {Dmitry}\ \bibnamefont
  {Budker}}\ and\ \bibinfo {author} {\bibfnamefont {Michael}\ \bibnamefont
  {Romalis}},\ }\bibfield  {title} {\enquote {\bibinfo {title} {Optical
  magnetometry},}\ }\href@noop {} {\bibfield  {journal} {\bibinfo  {journal}
  {Nature Physics}\ }\textbf {\bibinfo {volume} {3}},\ \bibinfo {pages} {227}
  (\bibinfo {year} {2007})}\BibitemShut {NoStop}%
\bibitem [{\citenamefont {Fleischhauer}\ \emph {et~al.}(2000)\citenamefont
  {Fleischhauer}, \citenamefont {Matsko},\ and\ \citenamefont
  {Scully}}]{Fleisch_magneto}%
  \BibitemOpen
  \bibfield  {author} {\bibinfo {author} {\bibfnamefont {M.}~\bibnamefont
  {Fleischhauer}}, \bibinfo {author} {\bibfnamefont {A.~B.}\ \bibnamefont
  {Matsko}}, \ and\ \bibinfo {author} {\bibfnamefont {M.~O.}\ \bibnamefont
  {Scully}},\ }\bibfield  {title} {\enquote {\bibinfo {title} {Quantum limit of
  optical magnetometry in the presence of ac stark shifts},}\ }\href {\doibase
  10.1103/PhysRevA.62.013808} {\bibfield  {journal} {\bibinfo  {journal} {Phys.
  Rev. A}\ }\textbf {\bibinfo {volume} {62}},\ \bibinfo {pages} {013808}
  (\bibinfo {year} {2000})}\BibitemShut {NoStop}%
\bibitem [{\citenamefont {Gerbier}\ \emph {et~al.}(2006)\citenamefont
  {Gerbier}, \citenamefont {Widera}, \citenamefont {F\"olling}, \citenamefont
  {Mandel},\ and\ \citenamefont {Bloch}}]{gerbier_pra_2006}%
  \BibitemOpen
  \bibfield  {author} {\bibinfo {author} {\bibfnamefont {Fabrice}\ \bibnamefont
  {Gerbier}}, \bibinfo {author} {\bibfnamefont {Artur}\ \bibnamefont {Widera}},
  \bibinfo {author} {\bibfnamefont {Simon}\ \bibnamefont {F\"olling}}, \bibinfo
  {author} {\bibfnamefont {Olaf}\ \bibnamefont {Mandel}}, \ and\ \bibinfo
  {author} {\bibfnamefont {Immanuel}\ \bibnamefont {Bloch}},\ }\bibfield
  {title} {\enquote {\bibinfo {title} {Resonant control of spin dynamics in
  ultracold quantum gases by microwave dressing},}\ }\href {\doibase
  10.1103/PhysRevA.73.041602} {\bibfield  {journal} {\bibinfo  {journal} {Phys.
  Rev. A}\ }\textbf {\bibinfo {volume} {73}},\ \bibinfo {pages} {041602}
  (\bibinfo {year} {2006})}\BibitemShut {NoStop}%
\bibitem [{\citenamefont {Jenkins}\ and\ \citenamefont
  {Ruostekoski}(2012{\natexlab{c}})}]{JenkinsLongPRB}%
  \BibitemOpen
  \bibfield  {author} {\bibinfo {author} {\bibfnamefont {S.~D.}\ \bibnamefont
  {Jenkins}}\ and\ \bibinfo {author} {\bibfnamefont {J.}~\bibnamefont
  {Ruostekoski}},\ }\bibfield  {title} {\enquote {\bibinfo {title} {Theoretical
  formalism for collective electromagnetic response of discrete metamaterial
  systems},}\ }\href@noop {} {\bibfield  {journal} {\bibinfo  {journal} {Phys.
  Rev. B}\ }\textbf {\bibinfo {volume} {86}},\ \bibinfo {pages} {085116}
  (\bibinfo {year} {2012}{\natexlab{c}})}\BibitemShut {NoStop}%
\bibitem [{\citenamefont {Ruostekoski}\ and\ \citenamefont
  {Javanainen}(1997)}]{Ruostekoski1997a}%
  \BibitemOpen
  \bibfield  {author} {\bibinfo {author} {\bibfnamefont {Janne}\ \bibnamefont
  {Ruostekoski}}\ and\ \bibinfo {author} {\bibfnamefont {Juha}\ \bibnamefont
  {Javanainen}},\ }\bibfield  {title} {\enquote {\bibinfo {title} {Quantum
  field theory of cooperative atom response: Low light intensity},}\
  }\href@noop {} {\bibfield  {journal} {\bibinfo  {journal} {Phys. Rev. A}\
  }\textbf {\bibinfo {volume} {55}},\ \bibinfo {pages} {513--526} (\bibinfo
  {year} {1997})}\BibitemShut {NoStop}%
\bibitem [{\citenamefont {Lee}\ \emph {et~al.}(2016)\citenamefont {Lee},
  \citenamefont {Jenkins},\ and\ \citenamefont {Ruostekoski}}]{Lee16}%
  \BibitemOpen
  \bibfield  {author} {\bibinfo {author} {\bibfnamefont {Mark~D.}\ \bibnamefont
  {Lee}}, \bibinfo {author} {\bibfnamefont {Stewart~D.}\ \bibnamefont
  {Jenkins}}, \ and\ \bibinfo {author} {\bibfnamefont {Janne}\ \bibnamefont
  {Ruostekoski}},\ }\bibfield  {title} {\enquote {\bibinfo {title} {Stochastic
  methods for light propagation and recurrent scattering in saturated and
  nonsaturated atomic ensembles},}\ }\href {\doibase
  10.1103/PhysRevA.93.063803} {\bibfield  {journal} {\bibinfo  {journal} {Phys.
  Rev. A}\ }\textbf {\bibinfo {volume} {93}},\ \bibinfo {pages} {063803}
  (\bibinfo {year} {2016})}\BibitemShut {NoStop}%
\bibitem [{\citenamefont {Jenkins}\ \emph
  {et~al.}(2016{\natexlab{a}})\citenamefont {Jenkins}, \citenamefont
  {Ruostekoski}, \citenamefont {Javanainen}, \citenamefont {Jennewein},
  \citenamefont {Bourgain}, \citenamefont {Pellegrino}, \citenamefont
  {Sortais},\ and\ \citenamefont {Browaeys}}]{Jenkins_long16}%
  \BibitemOpen
  \bibfield  {author} {\bibinfo {author} {\bibfnamefont {S.~D.}\ \bibnamefont
  {Jenkins}}, \bibinfo {author} {\bibfnamefont {J.}~\bibnamefont
  {Ruostekoski}}, \bibinfo {author} {\bibfnamefont {J.}~\bibnamefont
  {Javanainen}}, \bibinfo {author} {\bibfnamefont {S.}~\bibnamefont
  {Jennewein}}, \bibinfo {author} {\bibfnamefont {R.}~\bibnamefont {Bourgain}},
  \bibinfo {author} {\bibfnamefont {J.}~\bibnamefont {Pellegrino}}, \bibinfo
  {author} {\bibfnamefont {Y.~R.~P.}\ \bibnamefont {Sortais}}, \ and\ \bibinfo
  {author} {\bibfnamefont {A.}~\bibnamefont {Browaeys}},\ }\bibfield  {title}
  {\enquote {\bibinfo {title} {Collective resonance fluorescence in small and
  dense atom clouds: Comparison between theory and experiment},}\ }\href
  {\doibase 10.1103/PhysRevA.94.023842} {\bibfield  {journal} {\bibinfo
  {journal} {Phys. Rev. A}\ }\textbf {\bibinfo {volume} {94}},\ \bibinfo
  {pages} {023842} (\bibinfo {year} {2016}{\natexlab{a}})}\BibitemShut
  {NoStop}%
\bibitem [{\citenamefont {Sutherland}\ and\ \citenamefont
  {Robicheaux}(2017)}]{Sutherland_satur}%
  \BibitemOpen
  \bibfield  {author} {\bibinfo {author} {\bibfnamefont {R.~T.}\ \bibnamefont
  {Sutherland}}\ and\ \bibinfo {author} {\bibfnamefont {F.}~\bibnamefont
  {Robicheaux}},\ }\bibfield  {title} {\enquote {\bibinfo {title} {Degenerate
  zeeman ground states in the single-excitation regime},}\ }\href {\doibase
  10.1103/PhysRevA.96.053840} {\bibfield  {journal} {\bibinfo  {journal} {Phys.
  Rev. A}\ }\textbf {\bibinfo {volume} {96}},\ \bibinfo {pages} {053840}
  (\bibinfo {year} {2017})}\BibitemShut {NoStop}%
\bibitem [{\citenamefont {Jackson}(1999)}]{Jackson}%
  \BibitemOpen
  \bibfield  {author} {\bibinfo {author} {\bibfnamefont {John~David}\
  \bibnamefont {Jackson}},\ }\href@noop {} {\emph {\bibinfo {title} {Classical
  Electrodynamics}}},\ \bibinfo {edition} {3rd}\ ed.\ (\bibinfo  {publisher}
  {Wiley, New York},\ \bibinfo {year} {1999})\BibitemShut {NoStop}%
\bibitem [{\citenamefont {Javanainen}\ \emph {et~al.}(1999)\citenamefont
  {Javanainen}, \citenamefont {Ruostekoski}, \citenamefont {Vestergaard},\ and\
  \citenamefont {Francis}}]{Javanainen1999a}%
  \BibitemOpen
  \bibfield  {author} {\bibinfo {author} {\bibfnamefont {Juha}\ \bibnamefont
  {Javanainen}}, \bibinfo {author} {\bibfnamefont {Janne}\ \bibnamefont
  {Ruostekoski}}, \bibinfo {author} {\bibfnamefont {Bjarne}\ \bibnamefont
  {Vestergaard}}, \ and\ \bibinfo {author} {\bibfnamefont {Matthew~R.}\
  \bibnamefont {Francis}},\ }\bibfield  {title} {\enquote {\bibinfo {title}
  {One-dimensional modelling of light propagation in dense and degenerate
  samples},}\ }\href@noop {} {\bibfield  {journal} {\bibinfo  {journal} {Phys.
  Rev. A}\ }\textbf {\bibinfo {volume} {59}},\ \bibinfo {pages} {649--666}
  (\bibinfo {year} {1999})}\BibitemShut {NoStop}%
\bibitem [{\citenamefont {Javanainen}\ \emph {et~al.}(2014)\citenamefont
  {Javanainen}, \citenamefont {Ruostekoski}, \citenamefont {Li},\ and\
  \citenamefont {Yoo}}]{Javanainen2014a}%
  \BibitemOpen
  \bibfield  {author} {\bibinfo {author} {\bibfnamefont {Juha}\ \bibnamefont
  {Javanainen}}, \bibinfo {author} {\bibfnamefont {Janne}\ \bibnamefont
  {Ruostekoski}}, \bibinfo {author} {\bibfnamefont {Yi}~\bibnamefont {Li}}, \
  and\ \bibinfo {author} {\bibfnamefont {Sung-Mi}\ \bibnamefont {Yoo}},\
  }\bibfield  {title} {\enquote {\bibinfo {title} {Shifts of a resonance line
  in a dense atomic sample},}\ }\href {\doibase 10.1103/PhysRevLett.112.113603}
  {\bibfield  {journal} {\bibinfo  {journal} {Phys. Rev. Lett.}\ }\textbf
  {\bibinfo {volume} {112}},\ \bibinfo {pages} {113603} (\bibinfo {year}
  {2014})}\BibitemShut {NoStop}%
\bibitem [{\citenamefont {Javanainen}\ and\ \citenamefont
  {Ruostekoski}(2016)}]{JavanainenMFT}%
  \BibitemOpen
  \bibfield  {author} {\bibinfo {author} {\bibfnamefont {Juha}\ \bibnamefont
  {Javanainen}}\ and\ \bibinfo {author} {\bibfnamefont {Janne}\ \bibnamefont
  {Ruostekoski}},\ }\bibfield  {title} {\enquote {\bibinfo {title} {Light
  propagation beyond the mean-field theory of standard optics},}\ }\href
  {\doibase 10.1364/OE.24.000993} {\bibfield  {journal} {\bibinfo  {journal}
  {Opt. Express}\ }\textbf {\bibinfo {volume} {24}},\ \bibinfo {pages}
  {993--1001} (\bibinfo {year} {2016})}\BibitemShut {NoStop}%
\bibitem [{Note1()}]{Note1}%
  \BibitemOpen
  \bibinfo {note} {For the isotropic $J=0\rightarrow J'=1$ system $\protect
  \mathcal {H}$ is symmetric, and we can determine the biorthogonality
  condition $\protect \mathrm {v}_j^T \protect \mathrm {v}_i=\delta _{ji}$,
  except for possible zero-binorm states for which $\protect \mathrm {v}_j^T
  \protect \mathrm {v}_j=0$ (that we have not encountered in our
  system).}\BibitemShut {Stop}%
\bibitem [{\citenamefont {Alzar}\ \emph {et~al.}(2002)\citenamefont {Alzar},
  \citenamefont {Martinez},\ and\ \citenamefont {Nussenzveig}}]{eit_mech}%
  \BibitemOpen
  \bibfield  {author} {\bibinfo {author} {\bibfnamefont {C.~L.~Garrido}\
  \bibnamefont {Alzar}}, \bibinfo {author} {\bibfnamefont {M.~A.~G.}\
  \bibnamefont {Martinez}}, \ and\ \bibinfo {author} {\bibfnamefont
  {P.}~\bibnamefont {Nussenzveig}},\ }\bibfield  {title} {\enquote {\bibinfo
  {title} {Classical analog of electromagnetically induced transparency},}\
  }\href {\doibase 10.1119/1.1412644} {\bibfield  {journal} {\bibinfo
  {journal} {American Journal of Physics}\ }\textbf {\bibinfo {volume} {70}},\
  \bibinfo {pages} {37--41} (\bibinfo {year} {2002})}\BibitemShut {NoStop}%
\bibitem [{\citenamefont {Dutton}\ and\ \citenamefont {Hau}(2004)}]{DuttonHau}%
  \BibitemOpen
  \bibfield  {author} {\bibinfo {author} {\bibfnamefont {Zachary}\ \bibnamefont
  {Dutton}}\ and\ \bibinfo {author} {\bibfnamefont {Lene~Vestergaard}\
  \bibnamefont {Hau}},\ }\bibfield  {title} {\enquote {\bibinfo {title}
  {Storing and processing optical information with ultraslow light in
  bose-einstein condensates},}\ }\href {\doibase 10.1103/PhysRevA.70.053831}
  {\bibfield  {journal} {\bibinfo  {journal} {Phys. Rev. A}\ }\textbf {\bibinfo
  {volume} {70}},\ \bibinfo {pages} {053831} (\bibinfo {year}
  {2004})}\BibitemShut {NoStop}%
\bibitem [{\citenamefont {Dutton}\ and\ \citenamefont
  {Ruostekoski}(2004)}]{dutton_prl_2004}%
  \BibitemOpen
  \bibfield  {author} {\bibinfo {author} {\bibfnamefont {Zachary}\ \bibnamefont
  {Dutton}}\ and\ \bibinfo {author} {\bibfnamefont {Janne}\ \bibnamefont
  {Ruostekoski}},\ }\bibfield  {title} {\enquote {\bibinfo {title} {Transfer
  and storage of vortex states in light and matter waves},}\ }\href {\doibase
  10.1103/PhysRevLett.93.193602} {\bibfield  {journal} {\bibinfo  {journal}
  {Phys. Rev. Lett.}\ }\textbf {\bibinfo {volume} {93}},\ \bibinfo {pages}
  {193602} (\bibinfo {year} {2004})}\BibitemShut {NoStop}%
\bibitem [{\citenamefont {Liu}\ \emph {et~al.}(2001)\citenamefont {Liu},
  \citenamefont {Dutton}, \citenamefont {Behroozi},\ and\ \citenamefont
  {Hau}}]{LiuEtAlNature2001}%
  \BibitemOpen
  \bibfield  {author} {\bibinfo {author} {\bibfnamefont {C.}~\bibnamefont
  {Liu}}, \bibinfo {author} {\bibfnamefont {Z.}~\bibnamefont {Dutton}},
  \bibinfo {author} {\bibfnamefont {C.~H.}\ \bibnamefont {Behroozi}}, \ and\
  \bibinfo {author} {\bibfnamefont {L.~H.}\ \bibnamefont {Hau}},\ }\bibfield
  {title} {\enquote {\bibinfo {title} {Observation of coherent optical
  information storage in and atomic medium using halted light pulses},}\
  }\href@noop {} {\bibfield  {journal} {\bibinfo  {journal} {Nature}\ }\textbf
  {\bibinfo {volume} {409}},\ \bibinfo {pages} {490} (\bibinfo {year}
  {2001})}\BibitemShut {NoStop}%
\bibitem [{\citenamefont {Ruostekoski}\ \emph {et~al.}(2009)\citenamefont
  {Ruostekoski}, \citenamefont {Foot},\ and\ \citenamefont
  {Deb}}]{Ruostekoski09}%
  \BibitemOpen
  \bibfield  {author} {\bibinfo {author} {\bibfnamefont {J.}~\bibnamefont
  {Ruostekoski}}, \bibinfo {author} {\bibfnamefont {C.~J.}\ \bibnamefont
  {Foot}}, \ and\ \bibinfo {author} {\bibfnamefont {A.~B.}\ \bibnamefont
  {Deb}},\ }\bibfield  {title} {\enquote {\bibinfo {title} {Light scattering
  for thermometry of fermionic atoms in an optical lattice},}\ }\href {\doibase
  10.1103/PhysRevLett.103.170404} {\bibfield  {journal} {\bibinfo  {journal}
  {Phys. Rev. Lett.}\ }\textbf {\bibinfo {volume} {103}},\ \bibinfo {pages}
  {170404} (\bibinfo {year} {2009})}\BibitemShut {NoStop}%
\bibitem [{\citenamefont {Elliott}\ \emph {et~al.}(2015)\citenamefont
  {Elliott}, \citenamefont {Kozlowski}, \citenamefont {Caballero-Benitez},\
  and\ \citenamefont {Mekhov}}]{Elliottt15}%
  \BibitemOpen
  \bibfield  {author} {\bibinfo {author} {\bibfnamefont {T.~J.}\ \bibnamefont
  {Elliott}}, \bibinfo {author} {\bibfnamefont {W.}~\bibnamefont {Kozlowski}},
  \bibinfo {author} {\bibfnamefont {S.~F.}\ \bibnamefont {Caballero-Benitez}},
  \ and\ \bibinfo {author} {\bibfnamefont {I.~B.}\ \bibnamefont {Mekhov}},\
  }\bibfield  {title} {\enquote {\bibinfo {title} {Multipartite entangled
  spatial modes of ultracold atoms generated and controlled by quantum
  measurement},}\ }\href {\doibase 10.1103/PhysRevLett.114.113604} {\bibfield
  {journal} {\bibinfo  {journal} {Phys. Rev. Lett.}\ }\textbf {\bibinfo
  {volume} {114}},\ \bibinfo {pages} {113604} (\bibinfo {year}
  {2015})}\BibitemShut {NoStop}%
\bibitem [{Note2()}]{Note2}%
  \BibitemOpen
  \bibinfo {note} {In this limit we only have an equation for $ \protect
  \mathcal {P}_I$, since there is no coupling to $ \protect \mathcal {P}_P$ in
  Eq.~\protect \textup {\hbox {\mathsurround \z@ \protect \normalfont
  (\ignorespaces \ref {bothtwo}\unskip \@@italiccorr )}}.}\BibitemShut {Stop}%
\bibitem [{Note3()}]{Note3}%
  \BibitemOpen
  \bibinfo {note} {Owing to the different definition of the linewidth in
  Ref.~\cite {CAIT}, the atomic lattice result is obtained with the
  substitution $\Gamma _E\rightarrow \gamma /2$ in the expression of the
  meta-molecule linewidth.}\BibitemShut {Stop}%
\bibitem [{\citenamefont {Morsch}\ and\ \citenamefont
  {Oberthaler}(2006)}]{Morsch06}%
  \BibitemOpen
  \bibfield  {author} {\bibinfo {author} {\bibfnamefont {Oliver}\ \bibnamefont
  {Morsch}}\ and\ \bibinfo {author} {\bibfnamefont {Markus}\ \bibnamefont
  {Oberthaler}},\ }\bibfield  {title} {\enquote {\bibinfo {title} {Dynamics of
  bose-einstein condensates in optical lattices},}\ }\href {\doibase
  10.1103/RevModPhys.78.179} {\bibfield  {journal} {\bibinfo  {journal} {Rev.
  Mod. Phys.}\ }\textbf {\bibinfo {volume} {78}},\ \bibinfo {pages} {179--215}
  (\bibinfo {year} {2006})}\BibitemShut {NoStop}%
\bibitem [{\citenamefont {Gross}(2017)}]{com:gross}%
  \BibitemOpen
  \bibfield  {author} {\bibinfo {author} {\bibfnamefont {C.}~\bibnamefont
  {Gross}},\ }\href@noop {} {\enquote {\bibinfo {title} {Private
  communication},}\ } (\bibinfo {year} {2017})\BibitemShut {NoStop}%
\bibitem [{\citenamefont {Antezza}\ and\ \citenamefont
  {Castin}(2013)}]{Castin13}%
  \BibitemOpen
  \bibfield  {author} {\bibinfo {author} {\bibfnamefont {Mauro}\ \bibnamefont
  {Antezza}}\ and\ \bibinfo {author} {\bibfnamefont {Yvan}\ \bibnamefont
  {Castin}},\ }\bibfield  {title} {\enquote {\bibinfo {title} {Photonic band
  gap in an imperfect atomic diamond lattice: Penetration depth and effects of
  finite size and vacancies},}\ }\href {\doibase 10.1103/PhysRevA.88.033844}
  {\bibfield  {journal} {\bibinfo  {journal} {Phys. Rev. A}\ }\textbf {\bibinfo
  {volume} {88}},\ \bibinfo {pages} {033844} (\bibinfo {year}
  {2013})}\BibitemShut {NoStop}%
\bibitem [{\citenamefont {Ruostekoski}\ and\ \citenamefont
  {Javanainen}(2016)}]{Ruostekoski_waveguide}%
  \BibitemOpen
  \bibfield  {author} {\bibinfo {author} {\bibfnamefont {Janne}\ \bibnamefont
  {Ruostekoski}}\ and\ \bibinfo {author} {\bibfnamefont {Juha}\ \bibnamefont
  {Javanainen}},\ }\bibfield  {title} {\enquote {\bibinfo {title} {Emergence of
  correlated optics in one-dimensional waveguides for classical and quantum
  atomic gases},}\ }\href {\doibase 10.1103/PhysRevLett.117.143602} {\bibfield
  {journal} {\bibinfo  {journal} {Phys. Rev. Lett.}\ }\textbf {\bibinfo
  {volume} {117}},\ \bibinfo {pages} {143602} (\bibinfo {year}
  {2016})}\BibitemShut {NoStop}%
\bibitem [{\citenamefont {Gross}\ and\ \citenamefont
  {Haroche}(1982)}]{GrossHarochePhysRep1982}%
  \BibitemOpen
  \bibfield  {author} {\bibinfo {author} {\bibfnamefont {M.}~\bibnamefont
  {Gross}}\ and\ \bibinfo {author} {\bibfnamefont {S.}~\bibnamefont
  {Haroche}},\ }\bibfield  {title} {\enquote {\bibinfo {title} {Superradiance:
  An essay on the theory of collective spontaneous emission},}\ }\href@noop {}
  {\bibfield  {journal} {\bibinfo  {journal} {Phys. Rep.}\ }\textbf {\bibinfo
  {volume} {93}},\ \bibinfo {pages} {301} (\bibinfo {year} {1982})}\BibitemShut
  {NoStop}%
\bibitem [{\citenamefont {Goban}\ \emph {et~al.}(2015)\citenamefont {Goban},
  \citenamefont {Hung}, \citenamefont {Hood}, \citenamefont {Yu}, \citenamefont
  {Muniz}, \citenamefont {Painter},\ and\ \citenamefont
  {Kimble}}]{kimblesuper}%
  \BibitemOpen
  \bibfield  {author} {\bibinfo {author} {\bibfnamefont {A.}~\bibnamefont
  {Goban}}, \bibinfo {author} {\bibfnamefont {C.-L.}\ \bibnamefont {Hung}},
  \bibinfo {author} {\bibfnamefont {J.~D.}\ \bibnamefont {Hood}}, \bibinfo
  {author} {\bibfnamefont {S.-P.}\ \bibnamefont {Yu}}, \bibinfo {author}
  {\bibfnamefont {J.~A.}\ \bibnamefont {Muniz}}, \bibinfo {author}
  {\bibfnamefont {O.}~\bibnamefont {Painter}}, \ and\ \bibinfo {author}
  {\bibfnamefont {H.~J.}\ \bibnamefont {Kimble}},\ }\bibfield  {title}
  {\enquote {\bibinfo {title} {Superradiance for atoms trapped along a photonic
  crystal waveguide},}\ }\href {\doibase 10.1103/PhysRevLett.115.063601}
  {\bibfield  {journal} {\bibinfo  {journal} {Phys. Rev. Lett.}\ }\textbf
  {\bibinfo {volume} {115}},\ \bibinfo {pages} {063601} (\bibinfo {year}
  {2015})}\BibitemShut {NoStop}%
\bibitem [{\citenamefont {Solano}\ \emph {et~al.}(2017)\citenamefont {Solano},
  \citenamefont {Barberis-Blostein}, \citenamefont {Fatemi}, \citenamefont
  {Orozco},\ and\ \citenamefont {Rolston}}]{Solano_super}%
  \BibitemOpen
  \bibfield  {author} {\bibinfo {author} {\bibfnamefont {P.}~\bibnamefont
  {Solano}}, \bibinfo {author} {\bibfnamefont {P.}~\bibnamefont
  {Barberis-Blostein}}, \bibinfo {author} {\bibfnamefont {F.~K.}\ \bibnamefont
  {Fatemi}}, \bibinfo {author} {\bibfnamefont {L.~A.}\ \bibnamefont {Orozco}},
  \ and\ \bibinfo {author} {\bibfnamefont {S.~L.}\ \bibnamefont {Rolston}},\
  }\bibfield  {title} {\enquote {\bibinfo {title} {Super-radiance reveals
  infinite-range dipole interactions through a nanofiber},}\ }\href {\doibase
  10.1038/s41467-017-01994-3} {\bibfield  {journal} {\bibinfo  {journal}
  {Nature Communications}\ }\textbf {\bibinfo {volume} {8}},\ \bibinfo {pages}
  {1857} (\bibinfo {year} {2017})}\BibitemShut {NoStop}%
\bibitem [{\citenamefont {Roof}\ \emph {et~al.}(2016)\citenamefont {Roof},
  \citenamefont {Kemp}, \citenamefont {Havey},\ and\ \citenamefont
  {Sokolov}}]{Roof16}%
  \BibitemOpen
  \bibfield  {author} {\bibinfo {author} {\bibfnamefont {S.~J.}\ \bibnamefont
  {Roof}}, \bibinfo {author} {\bibfnamefont {K.~J.}\ \bibnamefont {Kemp}},
  \bibinfo {author} {\bibfnamefont {M.~D.}\ \bibnamefont {Havey}}, \ and\
  \bibinfo {author} {\bibfnamefont {I.~M.}\ \bibnamefont {Sokolov}},\
  }\bibfield  {title} {\enquote {\bibinfo {title} {Observation of single-photon
  superradiance and the cooperative lamb shift in an extended sample of cold
  atoms},}\ }\href {\doibase 10.1103/PhysRevLett.117.073003} {\bibfield
  {journal} {\bibinfo  {journal} {Phys. Rev. Lett.}\ }\textbf {\bibinfo
  {volume} {117}},\ \bibinfo {pages} {073003} (\bibinfo {year}
  {2016})}\BibitemShut {NoStop}%
\bibitem [{\citenamefont {Ara\'ujo}\ \emph {et~al.}(2016)\citenamefont
  {Ara\'ujo}, \citenamefont {Kre\ifmmode \check{s}\else
  \v{s}\fi{}i\ifmmode~\acute{c}\else \'{c}\fi{}}, \citenamefont {Kaiser},\ and\
  \citenamefont {Guerin}}]{Araujo16}%
  \BibitemOpen
  \bibfield  {author} {\bibinfo {author} {\bibfnamefont {Michelle~O.}\
  \bibnamefont {Ara\'ujo}}, \bibinfo {author} {\bibfnamefont {Ivor}\
  \bibnamefont {Kre\ifmmode \check{s}\else \v{s}\fi{}i\ifmmode~\acute{c}\else
  \'{c}\fi{}}}, \bibinfo {author} {\bibfnamefont {Robin}\ \bibnamefont
  {Kaiser}}, \ and\ \bibinfo {author} {\bibfnamefont {William}\ \bibnamefont
  {Guerin}},\ }\bibfield  {title} {\enquote {\bibinfo {title} {Superradiance in
  a large and dilute cloud of cold atoms in the linear-optics regime},}\ }\href
  {\doibase 10.1103/PhysRevLett.117.073002} {\bibfield  {journal} {\bibinfo
  {journal} {Phys. Rev. Lett.}\ }\textbf {\bibinfo {volume} {117}},\ \bibinfo
  {pages} {073002} (\bibinfo {year} {2016})}\BibitemShut {NoStop}%
\bibitem [{\citenamefont {Kwong}\ \emph {et~al.}(2014)\citenamefont {Kwong},
  \citenamefont {Yang}, \citenamefont {Pramod}, \citenamefont {Pandey},
  \citenamefont {Delande}, \citenamefont {Pierrat},\ and\ \citenamefont
  {Wilkowski}}]{wilkowski}%
  \BibitemOpen
  \bibfield  {author} {\bibinfo {author} {\bibfnamefont {C.~C.}\ \bibnamefont
  {Kwong}}, \bibinfo {author} {\bibfnamefont {T.}~\bibnamefont {Yang}},
  \bibinfo {author} {\bibfnamefont {M.~S.}\ \bibnamefont {Pramod}}, \bibinfo
  {author} {\bibfnamefont {K.}~\bibnamefont {Pandey}}, \bibinfo {author}
  {\bibfnamefont {D.}~\bibnamefont {Delande}}, \bibinfo {author} {\bibfnamefont
  {R.}~\bibnamefont {Pierrat}}, \ and\ \bibinfo {author} {\bibfnamefont
  {D.}~\bibnamefont {Wilkowski}},\ }\bibfield  {title} {\enquote {\bibinfo
  {title} {Cooperative emission of a coherent superflash of light},}\ }\href
  {\doibase 10.1103/PhysRevLett.113.223601} {\bibfield  {journal} {\bibinfo
  {journal} {Phys. Rev. Lett.}\ }\textbf {\bibinfo {volume} {113}},\ \bibinfo
  {pages} {223601} (\bibinfo {year} {2014})}\BibitemShut {NoStop}%
\bibitem [{\citenamefont {Keaveney}\ \emph {et~al.}(2012)\citenamefont
  {Keaveney}, \citenamefont {Sargsyan}, \citenamefont {Krohn}, \citenamefont
  {Hughes}, \citenamefont {Sarkisyan},\ and\ \citenamefont
  {Adams}}]{Keaveney2012}%
  \BibitemOpen
  \bibfield  {author} {\bibinfo {author} {\bibfnamefont {J.}~\bibnamefont
  {Keaveney}}, \bibinfo {author} {\bibfnamefont {A.}~\bibnamefont {Sargsyan}},
  \bibinfo {author} {\bibfnamefont {U.}~\bibnamefont {Krohn}}, \bibinfo
  {author} {\bibfnamefont {I.~G.}\ \bibnamefont {Hughes}}, \bibinfo {author}
  {\bibfnamefont {D.}~\bibnamefont {Sarkisyan}}, \ and\ \bibinfo {author}
  {\bibfnamefont {C.~S.}\ \bibnamefont {Adams}},\ }\bibfield  {title} {\enquote
  {\bibinfo {title} {Cooperative {Lamb} shift in an atomic vapor layer of
  nanometer thickness},}\ }\href@noop {} {\bibfield  {journal} {\bibinfo
  {journal} {Phys. Rev. Lett.}\ }\textbf {\bibinfo {volume} {108}},\ \bibinfo
  {pages} {173601} (\bibinfo {year} {2012})}\BibitemShut {NoStop}%
\bibitem [{\citenamefont {Meir}\ \emph {et~al.}(2014)\citenamefont {Meir},
  \citenamefont {Schwartz}, \citenamefont {Shahmoon}, \citenamefont {Oron},\
  and\ \citenamefont {Ozeri}}]{Meir13}%
  \BibitemOpen
  \bibfield  {author} {\bibinfo {author} {\bibfnamefont {Z.}~\bibnamefont
  {Meir}}, \bibinfo {author} {\bibfnamefont {O.}~\bibnamefont {Schwartz}},
  \bibinfo {author} {\bibfnamefont {E.}~\bibnamefont {Shahmoon}}, \bibinfo
  {author} {\bibfnamefont {D.}~\bibnamefont {Oron}}, \ and\ \bibinfo {author}
  {\bibfnamefont {R.}~\bibnamefont {Ozeri}},\ }\bibfield  {title} {\enquote
  {\bibinfo {title} {Cooperative lamb shift in a mesoscopic atomic array},}\
  }\href {\doibase 10.1103/PhysRevLett.113.193002} {\bibfield  {journal}
  {\bibinfo  {journal} {Phys. Rev. Lett.}\ }\textbf {\bibinfo {volume} {113}},\
  \bibinfo {pages} {193002} (\bibinfo {year} {2014})}\BibitemShut {NoStop}%
\bibitem [{\citenamefont {R\"ohlsberger}\ \emph {et~al.}(2010)\citenamefont
  {R\"ohlsberger}, \citenamefont {Schlage}, \citenamefont {Sahoo},
  \citenamefont {Couet},\ and\ \citenamefont {R\"uffer}}]{ROH10}%
  \BibitemOpen
  \bibfield  {author} {\bibinfo {author} {\bibfnamefont {Ralf}\ \bibnamefont
  {R\"ohlsberger}}, \bibinfo {author} {\bibfnamefont {Kai}\ \bibnamefont
  {Schlage}}, \bibinfo {author} {\bibfnamefont {Balaram}\ \bibnamefont
  {Sahoo}}, \bibinfo {author} {\bibfnamefont {Sebastien}\ \bibnamefont
  {Couet}}, \ and\ \bibinfo {author} {\bibfnamefont {Rudolf}\ \bibnamefont
  {R\"uffer}},\ }\bibfield  {title} {\enquote {\bibinfo {title} {Collective
  {L}amb shift in single-photon superradiance},}\ }\href {\doibase
  10.1126/science.1187770} {\bibfield  {journal} {\bibinfo  {journal}
  {Science}\ }\textbf {\bibinfo {volume} {328}},\ \bibinfo {pages} {1248--1251}
  (\bibinfo {year} {2010})}\BibitemShut {NoStop}%
\bibitem [{\citenamefont {Jenkins}\ \emph
  {et~al.}(2016{\natexlab{b}})\citenamefont {Jenkins}, \citenamefont
  {Ruostekoski}, \citenamefont {Javanainen}, \citenamefont {Bourgain},
  \citenamefont {Jennewein}, \citenamefont {Sortais},\ and\ \citenamefont
  {Browaeys}}]{Jenkins_thermshift}%
  \BibitemOpen
  \bibfield  {author} {\bibinfo {author} {\bibfnamefont {S.~D.}\ \bibnamefont
  {Jenkins}}, \bibinfo {author} {\bibfnamefont {J.}~\bibnamefont
  {Ruostekoski}}, \bibinfo {author} {\bibfnamefont {J.}~\bibnamefont
  {Javanainen}}, \bibinfo {author} {\bibfnamefont {R.}~\bibnamefont
  {Bourgain}}, \bibinfo {author} {\bibfnamefont {S.}~\bibnamefont {Jennewein}},
  \bibinfo {author} {\bibfnamefont {Y.~R.~P.}\ \bibnamefont {Sortais}}, \ and\
  \bibinfo {author} {\bibfnamefont {A.}~\bibnamefont {Browaeys}},\ }\bibfield
  {title} {\enquote {\bibinfo {title} {Optical resonance shifts in the
  fluorescence of thermal and cold atomic gases},}\ }\href {\doibase
  10.1103/PhysRevLett.116.183601} {\bibfield  {journal} {\bibinfo  {journal}
  {Phys. Rev. Lett.}\ }\textbf {\bibinfo {volume} {116}},\ \bibinfo {pages}
  {183601} (\bibinfo {year} {2016}{\natexlab{b}})}\BibitemShut {NoStop}%
\bibitem [{\citenamefont {Jennewein}\ \emph {et~al.}(2016)\citenamefont
  {Jennewein}, \citenamefont {Besbes}, \citenamefont {Schilder}, \citenamefont
  {Jenkins}, \citenamefont {Sauvan}, \citenamefont {Ruostekoski}, \citenamefont
  {Greffet}, \citenamefont {Sortais},\ and\ \citenamefont
  {Browaeys}}]{Jennewein_trans}%
  \BibitemOpen
  \bibfield  {author} {\bibinfo {author} {\bibfnamefont {S.}~\bibnamefont
  {Jennewein}}, \bibinfo {author} {\bibfnamefont {M.}~\bibnamefont {Besbes}},
  \bibinfo {author} {\bibfnamefont {N.~J.}\ \bibnamefont {Schilder}}, \bibinfo
  {author} {\bibfnamefont {S.~D.}\ \bibnamefont {Jenkins}}, \bibinfo {author}
  {\bibfnamefont {C.}~\bibnamefont {Sauvan}}, \bibinfo {author} {\bibfnamefont
  {J.}~\bibnamefont {Ruostekoski}}, \bibinfo {author} {\bibfnamefont {J.-J.}\
  \bibnamefont {Greffet}}, \bibinfo {author} {\bibfnamefont {Y.~R.~P.}\
  \bibnamefont {Sortais}}, \ and\ \bibinfo {author} {\bibfnamefont
  {A.}~\bibnamefont {Browaeys}},\ }\bibfield  {title} {\enquote {\bibinfo
  {title} {Coherent scattering of near-resonant light by a dense microscopic
  cold atomic cloud},}\ }\href {\doibase 10.1103/PhysRevLett.116.233601}
  {\bibfield  {journal} {\bibinfo  {journal} {Phys. Rev. Lett.}\ }\textbf
  {\bibinfo {volume} {116}},\ \bibinfo {pages} {233601} (\bibinfo {year}
  {2016})}\BibitemShut {NoStop}%
\bibitem [{\citenamefont {Corman}\ \emph {et~al.}(2017)\citenamefont {Corman},
  \citenamefont {Ville}, \citenamefont {Saint-Jalm}, \citenamefont
  {Aidelsburger}, \citenamefont {Bienaim\'e}, \citenamefont {Nascimb\`ene},
  \citenamefont {Dalibard},\ and\ \citenamefont {Beugnon}}]{Dalibard_slab}%
  \BibitemOpen
  \bibfield  {author} {\bibinfo {author} {\bibfnamefont {L.}~\bibnamefont
  {Corman}}, \bibinfo {author} {\bibfnamefont {J.~L.}\ \bibnamefont {Ville}},
  \bibinfo {author} {\bibfnamefont {R.}~\bibnamefont {Saint-Jalm}}, \bibinfo
  {author} {\bibfnamefont {M.}~\bibnamefont {Aidelsburger}}, \bibinfo {author}
  {\bibfnamefont {T.}~\bibnamefont {Bienaim\'e}}, \bibinfo {author}
  {\bibfnamefont {S.}~\bibnamefont {Nascimb\`ene}}, \bibinfo {author}
  {\bibfnamefont {J.}~\bibnamefont {Dalibard}}, \ and\ \bibinfo {author}
  {\bibfnamefont {J.}~\bibnamefont {Beugnon}},\ }\bibfield  {title} {\enquote
  {\bibinfo {title} {Transmission of near-resonant light through a dense slab
  of cold atoms},}\ }\href {\doibase 10.1103/PhysRevA.96.053629} {\bibfield
  {journal} {\bibinfo  {journal} {Phys. Rev. A}\ }\textbf {\bibinfo {volume}
  {96}},\ \bibinfo {pages} {053629} (\bibinfo {year} {2017})}\BibitemShut
  {NoStop}%
\end{thebibliography}
\end{document}